\documentclass[useAMS,usenatbib,usegraphicx]{mn2e}

\usepackage{hyperref} \usepackage{color} \usepackage{verbatim}
\usepackage{longtable} \usepackage{threeparttable}
\usepackage{multicol} \usepackage{lscape} \usepackage{subfig}
\usepackage{gensymb} \usepackage{float} 

\title[Searching for nova shells around cataclysmic variables -
  II]{Searching for nova shells around cataclysmic variables - II. A
  second campaign.}  \author[D. I. Sahman et
  al]{D. I. Sahman,$^{1}$\thanks{E-mail: david.sahman@sheffield.ac.uk}
  V. S. Dhillon$^{1,2}$. \\$^{1}$Department of Physics and Astronomy,
  University of Sheffield, Sheffield S3 7RH, UK\\ $^{2}$Instituto de
  Astrof\'{i}sica de Canarias, E-38205 La Laguna, Tenerife, Spain\\}

\begin{document}

\date{Accepted 2021 December 13. Received 2021 November 13}


\maketitle

\label{firstpage}

\begin{abstract}

 We report on our second campaign to search for old nova shells around
 cataclysmic variables (CVs). Our aim was to test the theory that nova eruptions
 cause cycles in the mass transfer rates of CVs. These mass transfer cycles
 change the behaviour of CVs during their inter-eruption periods. We examined H$
 \alpha$ images of 47 objects and found no new shells around any of the targets.
 Combining our latest results with our previous campaign \citep{sahman15}, and
 the searches by \citet{schmidtobreick15} and \citet{pagnotta16}, we estimate
 that the nova-like phase of the mass transfer cycle lasts $\sim$3,000 years.

\end{abstract}


\section{Introduction}
 
 Cataclysmic variables (CVs) are close binary systems in which a white dwarf
 (WD) primary accretes material from a late-type secondary star via Roche-lobe
 overflow (see \citealt{warner95a} for a review). Non-magnetic CVs are
 classified into 3 main sub-types -- the classical novae, the dwarf novae and
 the nova-likes. The {\em classical novae} (CNe) are defined as systems in which
 only a single nova eruption has been observed. Nova eruptions have typical
 amplitudes of 10 magnitudes and are believed to be due to the thermonuclear
 runaway of hydrogen-rich material accreted onto the surface of the white dwarf.
 The {\em dwarf novae} (DNe) are defined as systems which undergo quasi-regular
 (on timescales of weeks--months) outbursts of much smaller amplitude (typically
 6 magnitudes). Dwarf novae outbursts are believed to be due to instabilities in
 the accretion disc causing it to collapse onto the white dwarf \citep{osaki74}.
 The {\em nova-like} variables (NLs) are the non-eruptive CVs, i.e. objects
 which have never been observed to show novae or dwarf novae outbursts. The
 absence of dwarf novae outbursts in NLs is believed to be due to their high
 mass-transfer rates, producing ionised accretion discs in which the
 disc-instability mechanism that causes DNe outbursts is suppressed; the mass
 transfer rates in NLs are $\dot{M} \sim 10^{-9}$ M$_{\odot}$ yr$^{-1}$ whereas
 DNe have rates of $\dot{M} \sim 10^{-11}$ M$_{\odot}$ yr$^{-1}$
 \citep{warner95a}.

 Mass transfer is believed to be driven by angular momentum loss, which in turn
 is driven by two main processes, mass loss from magnetically-coupled winds from
 the secondary star (commonly referred to as \textit{magnetic braking}), and
 gravitational wave emission. At periods longer than about 3 hours, both
 mechanisms operate, leading to high $ \dot{M}$ as seen in NLs. At orbital
 periods below two hours, only gravitational wave emission occurs, and $\dot{M}$
 is lower, as in DNe. However, the orbital period distribution of CVs shows both
 high and low $\dot{M}$ systems co-existing at the same orbital periods
 \citep{knigge11}.

One explanation for the coexistence of systems at the same orbital period with
high and low $\dot{M}$ is a nova-induced cycle. Some fraction of the energy
released in the nova event will heat up the WD, leading to irradiation and
subsequent bloating of the secondary. Following the nova event, the system would
have a high $\dot{M}$ and appear as a NL. As the WD cools, $\dot{M}$ reduces and
the system changes to a DN, or even possibly $\dot{M}$ ceases altogether and the
system goes into hibernation.  Hence CVs are expected to cycle between nova, NL
and DN states, on timescales of $10^4-10^5$ yrs (\citealt{shara86},
\citealt{bode08}). This nova-induced cycle theory became known as hibernation
theory

Hibernation theory was originally invoked to explain the disparity between the
observed space density of CVs compared to models. Recent surveys using Gaia DR2
data \citep{pala19} compared to more sophisticated models \citep{belloni18} have
shown that the space density of CVs is broadly in agreement with theoretical
predictions. However, we still need to understand the impact of the nova
eruption on the evolution of CVs. Recent modelling by \citet{hillman20} predicts
that some systems are expected to cease mass transfer after a nova eruption and
go into hibernation, though far fewer systems that exhibit this behaviour are
predicted than were first proposed in the original version of the theory
\citep{shara86}. Recently however, \citet{schaefer20} reported measurements the
orbital period changes of six of CNe, using a long term monitoring campaign and
archival data. He found that five of them showed a period decrease  across the
nova eruption. This is exactly the opposite of the model predictions, which
suggest that the orbital period should increase as the two stars are driven
apart during the nova eruption. 

The cyclical evolution of CVs through CN, NL and DN phases has received
observational support from the discovery that BK Lyn appears to have evolved
through all three phases since its likely nova outburst in the year AD 101
\citep{patterson13}. A second piece of evidence has come from the discovery of
nova shells associated with the dwarf novae Z Cam, AT Cnc and Nova Sco 1437
(\citealt{shara07}, \citealt{shara13}, \citealt{shara17}). However, the
association of these CVs with the historical Chinese and Korean records has been
reviewed by \citet{hoffmann19}, who found that all three are doubtful.

A more obvious place than DNe to find nova shells is around NLs, as the nova-
induced cycle theory suggests that the high $\dot{M}$ in NLs is due to a recent
nova outburst. This was the motivation behind our first campaign to search for
nova shells around CVs, reported in \citeauthor{sahman15} (2015, hereafter S15).
We discovered a nebula around the nova-like V1315 Aql, which was subsequently
confirmed as a nova shell using Keck DEIMOS spectroscopy \citep{sahman18}.
Other shells have since been discovered around other nova-likes. IPHASX
J210204.7+471015 is a NL at the centre of a shell originally thought to be a
planetary nebula \citep{guerrero18}. V341 Ara is another NL at the centre of a
shell originally thought to be a planetary nebula \citep{castro20}. BZ Cam is
located in a bow-shock nebula with faint circular filaments, and is possibly
associated with the ancient nova in AD 369 \citep{hoffmann20}.

Our first campaign, which relied on H$ \alpha$ images obtained with both the
Auxiliary Port Camera on the 4.2m William Herschel Telescope (WHT) and the 2.5m
Isaac Newton Telescope (INT) Photometric H$ \alpha$ Survey of the Northern
Galactic Plane (IPHAS), suffered from two drawbacks. First, the field of view of
the WHT was too small to detect the large shells which have been discovered
around DNe (\citealt{shara07}, \citealt{shara13}), and also contained too few
field stars to derive a reliable stellar point spread function (PSF) which
allows one to detect shells close to the CVs. The second problem was that the
wide--field IPHAS survey was too shallow, and was targeted at the Galactic plane
which has large background H$\alpha$ nebulosity, making it difficult to detect
faint nova shells.

In this paper, we report on our second, much deeper and wider--field campaign to
identify old nova shells around CVs, and use our results, combined with other
searches, to estimate the lifetime of the nova-like phase following a nova
eruption.

\section{Target selection, Observations, and Data Reduction}

\subsection{Target selection}
\label{inttarget}

This second campaign was motivated by the shortcomings of our original campaign
in S15. We had estimated the age of the faintest shell we could detect with the
WHT by simulating images using the old nova DQ~Her. The simulations showed that
we could expect to detect shells at $\sim180$ yr after outburst. This detection
threshold can be pushed fainter, and hence older by $\sim{40}$ years, by
computing the mean radial profile of the PSF of the central object and
inspecting the wings for evidence of a shell, a technique pioneered by
\citet{gill98}.

However, using DQ~Her led us to underestimate the optimal field of view for
hunting for nova shells.  This is because DQ~Her has relatively slow-moving
ejecta (350\,km\,s$^{-1}$; \citealt{warner95a}). The angular size of a nova
shell is determined by the time since the nova eruption, the distance to the CV,
and the speed of the ejecta, and is given by the following scaling relation:

\begin{equation}
 \qquad \qquad \qquad R\sim 20\arcsec \,\, \frac{ t/100\,\mathrm{yr}
   \times v/1000\,\mathrm{km \,s}^{-1}}{d/\mathrm{kpc}}, 
     \end{equation}
where $R$ is the angular radius of the shell in arcseconds, $t$ is the time
elapsed since the nova eruption, $v$ is the shell expansion velocity, and $d$ is
the distance to the CV. Hence a recent, distant nova with slow-moving ejecta
($t=100\,$yr$, v=500\,$km\,s$^{-1}, d=2\,$kpc) would have a small shell of
radius $\sim5\arcsec$, whereas an older, nearby nova with fast moving ejecta
($t=200\,$yr$, v=2000\,$km\,s$^{-1}, d=0.5\,$kpc) would have expanded to a
radius of $\sim 2.7\arcmin$.  The field of view of the Auxiliary Port Camera on
the WHT ($\sim1\arcmin$ radius) used for S15, was too small to detect such large
shells. This is borne out by the size of the shell that we discovered around
V1315 Aql \citep{sahman18}, which is $\sim2.5\arcmin$ in radius, the shell
around IPHASX J210204.7+471015 which has a radius of $\sim 2.0\arcmin$
\citep{guerrero18}, and the shells discovered by \citeauthor{shara07} (2007,
2012) of radii $1.5\arcmin$ (AT~Cnc) and $15\arcmin$ (Z~Cam). A further problem
with having such a small field of view is the paucity of field stars that are
available to compare radial profiles (see Section \ref{im}).

Whilst the IPHAS survey had a sufficiently large field of view ($\sim17\arcmin$
radius per pointing) to discover nova shells, it suffered from very short
exposure times (120\,s). Furthermore, IPHAS was constrained to the Galactic
plane, making it difficult to pick out faint nova shells from the bright
background H$\alpha$ nebulosity. We concluded that a more optimal survey for
nova shells would have approximately the same field of view as an IPHAS
pointing, avoid the Galactic plane and be of similar depth to our WHT survey
($\sim{25}$~mags/arcsec$^2$).

To ensure we only targeted relatively well-studied systems with reliable periods
and CV classifications, we made our selection from version 7.20 of the catalogue
of \citet{ritter03}\footnote{http://www.mpa-garching.mpg.de/RKcat/} (hereafter
RK catalogue). We deliberately selected a substantial number of the systems in
the 3--4\,hr orbital period range, which is where most NLs appear, as shown in
Fig.~18 of \citet{knigge11}.

We included a number of asynchronous polars in the target list. Asynchronous
polars are CVs with a highly magnetic WD, in which the WD spin period is not
synchronous with the orbital period. One possible cause of the asynchronicity is
believed to be a nova eruption \citep{campbell99}. The target list also included
a few DNe as a control sample and to ensure that we had sufficient targets
visible on each night. All targets were outside the central galactic plane, with
a galactic latitude \textit{b} $> \pm 5^{o}$.

\subsection{Observations}
\label{obs}

We were allocated three nights on the INT in August 2014 and three nights in
January 2015.  We were also allocated time from the service observing programme
which resulted in an additional 10 targets being observed. We observed 47
targets in total, primarily NLs, including 16 systems that we had previously
examined in S15 using the WHT. The targets comprised one old nova (which is also
an intermediate polar), three asynchronous polars, 39 NLs, and four DNe. 

We used the WFC mounted at the prime focus of the INT. The WFC consists of 4
thinned e2v 2kx4k CCDs and has a platescale of 0.33$\arcsec$ per pixel, with a
field of view of $\sim$~34$\arcmin$ x 34$\arcmin$. In order to cover the gaps in
the CCD array, we took sets of four images dithered by 20\arcsec \,up and down,
and right and left.

We used the H$\alpha$ filter no. 197, which is centred on H$\alpha$ and has a
FWHM of 95\AA. Note that this filter also includes a contribution from N[{\sc
ii}] 6584\AA \,emission, which may dominate the spectra of nova shells with
strong shock interaction of the ejecta with any pre-existing circumstellar
medium, e.g. T Pyx \citep{shara89}. We also used the Sloan $r$ band filter no.
214 (centred on 6240\AA \,with FWHM 1347\AA).

A full list of the 47 objects and a journal of observations is given in
Table1~\ref{tab:journal}. The observing conditions were generally good, with
seeing below $2$\arcsec. There was some cloud on the nights of 15 December 2014,
and 22 February 2015, and on those nights the seeing also worsened to 3\arcsec\,
and 4\arcsec\,, respectively.

\subsection{Data reduction}

The images were processed using the data reduction pipeline THELI
\citep{schirmer13}.  Bias correction was carried out using bias frames taken at
the start and the end of the night.  Flat-fielding was performed using twilight
flats. The astrometry was performed using SCAMP \citep{bertin06}, with standard
astrometric catalogues (eg. PPMXL, USNO, 2MASS). The processed images were
co--added by taking the median. No sky subtraction was performed to avoid the
accidental removal of faint nova shells.

\section{Results}
\label{sec:res}

\subsection{INT Images and Radial Profiles}
\label{im}

\begin{figure}
\centering
  \vspace{10pt} \includegraphics[width=80mm,angle=0]{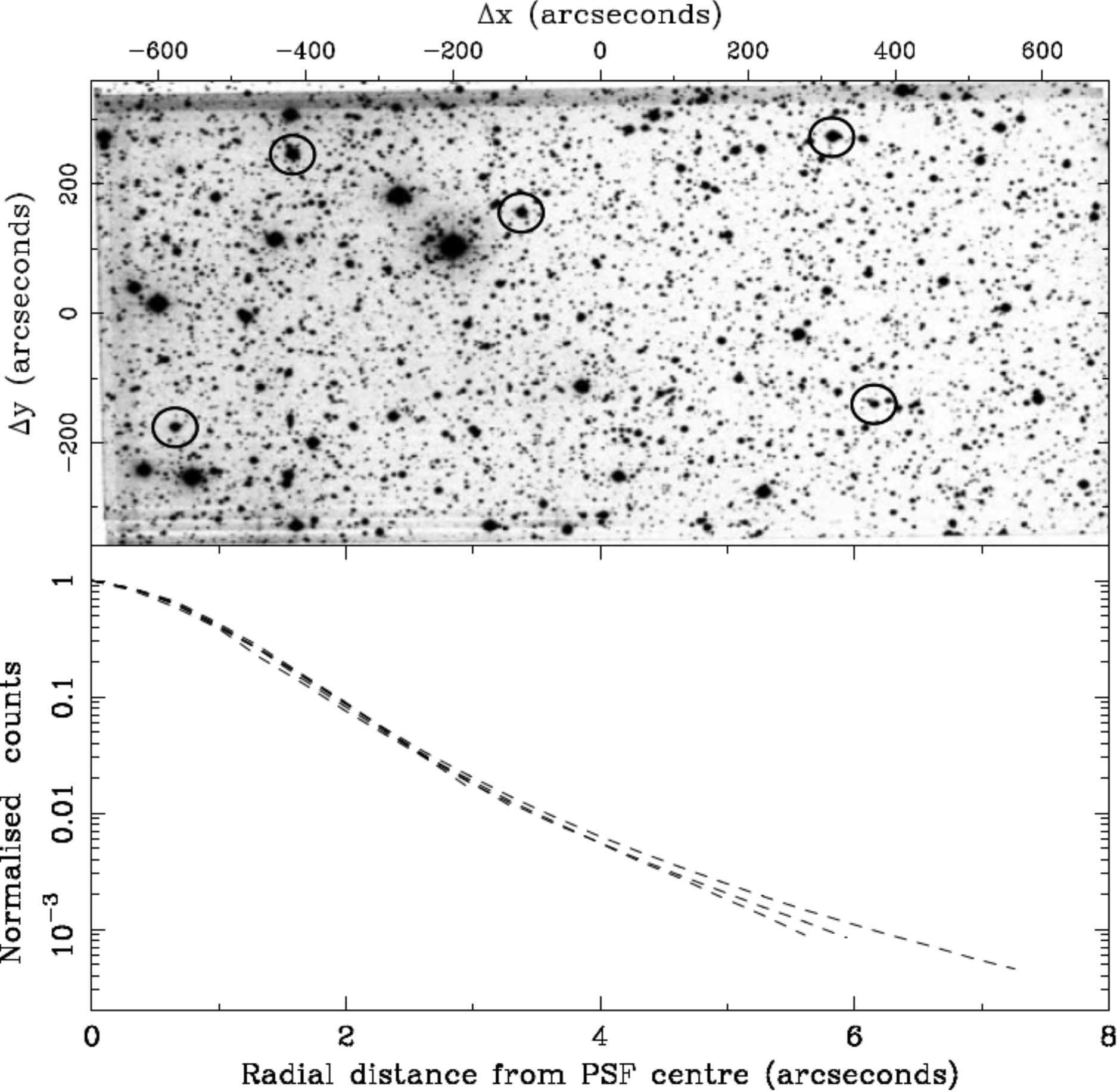}
\caption{PSF Test for one of the CCDs in the WFC. The PSFs of the circled stars
in the upper frame are plotted in the lower frame. Due to nearby stars, the
maximum radial distance plotted varies for each star.}
\label{fig:psftest1}
\end{figure}

In order to detect shells in the images, we adopted two strategies. First, we
visually examined each image to determine if a shell is visible. This technique
reveals shells with diameters of more than a few arcseconds. Second, we
calculated the radial profile (PSF) of each CV and compared it to the radial
profiles of a number of field stars in the same image. Any nebulosity around the
CV due to a nova shell would cause the radial profile of the CV to lie above the
average profile of the field stars. This technique can reveal shells with
diameters of less than a few arcseconds, and was successfully used by
\citet{gill98} to discover four new nova shells. We also used this technique in
S15. A key assumption in this technique is that the radial profile of the PSF is
uniform across the field of view. Fig \ref{fig:psftest1} shows the PSFs for
various field stars (circled) in the image of V1432 Aql. The PSFs show identical
radial profiles irrespective of field position, giving confidence that the PSFs
are uniform across the field of view of the CCD, as expected.

The images and radial profiles of our 47 targets are shown in Appendix A, except
for the image of V1315 Aql which is presented in \citet{sahman18}. Note that
the contrast in the images has been optimised. The background sky level appears
white and any emission brighter than a few percent above the background,
including the stars, appears black. This emphasises any faint nebular emission.
As a result, the background levels of each of the four chips appears slightly
different in some of the images.

\begin{figure}
\centering
  \vspace{10pt} \includegraphics[width=80mm,angle=0]{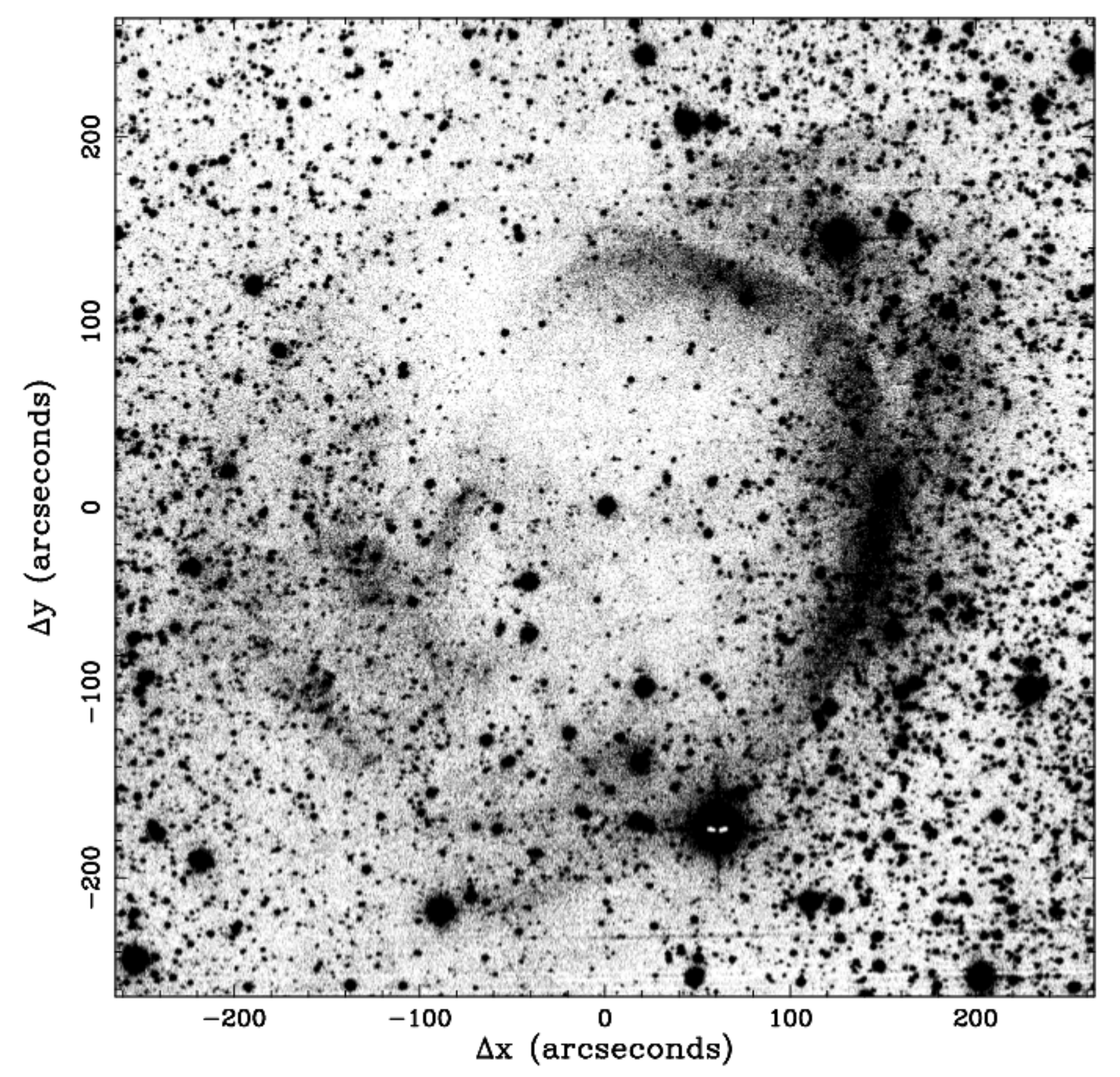}
\caption{INT WFC H$\alpha$ image of the nova shell around V1315
  Aql. The binary is located at the centre of the image. North is up
  and East is left.}
\label{fig:v1315aql}
\end{figure}

 \begin{table*}
   \begin{minipage}{160mm}
     \begin{center}
\caption[]{Journal of observations (in alphabetical order of
  constellation). The classifications and orbital periods of the CVs
  have been taken from the RK catalogue \citep{ritter03}. The date
  refers to the start time of the first exposure.}
\begin{tabular}{llccccccc}
  \hline \hline & & & & & & & & \\ 
 \multicolumn{1}{l}{Object} &
\multicolumn{1}{l}{Classification} & \multicolumn{1}{c}{Orbital} &
\multicolumn{1}{c}{Date} & \multicolumn{1}{c}{UTC} &
\multicolumn{1}{c}{UTC} & \multicolumn{1}{c}{Number of} &
\multicolumn{1}{c}{Total exposure} & \multicolumn{1}{c}{Visible} \\ &
& \multicolumn{1}{c}{period (hrs)} & & \multicolumn{1}{c}{start} &
\multicolumn{1}{c}{end} & \multicolumn{1}{c}{exposures} &
 \multicolumn{1}{c}{time (secs)} & \multicolumn{1}{c}{shell?} \\
 & & & & & & & & \\ \hline
PX And & NL SW NS SH & 3.51 & 01/08/14 & 04:43 & 05:16 &
2 & 1800 & N \\ & & & 04/08/14 & 03:18 & 03:42 & 4 & 1080 & N \\ HL
Aqr & NL UX SW & 3.25 & 04/08/14 & 00:27 & 02:10 & 8 & 3840 & N \\ UU
Aqr & NL UX SW SH & 3.93 & 01/08/14 & 03:16 & 04:36 & 7 & 3780 & N
\\ V794 Aql & NL VY & 3.68 & 03/08/14 & 22:39 & 23:37 & 6 & 2880 & N
\\ V1315 Aql & NL UX SW & 3.35 & 01/08/14 & 22:45 & 01:55 & 13 & 8700
& Y \\ V1432 Aql & NL AM AS & 3.37 & 02/08/14 & 21:43 & 23:16 & 32 &
3840 & N \\ WX Ari & NL UX SW & 3.34 & 15/01/15 & 20:37 & 22:01 & 8 &
3840 & N\\ KR Aur & NL VY NS & 3.91 & 18/01/15 & 01:26 & 02:45 & 8 &
3840 & N \\ BY Cam & NL AM AS & 3.35 & 15/01/15 & 23:48 & 01:26 & 8 &
3840 & N \\ BZ Cam & NL VY SH & 3.69 & 16/12/14 & 05:19 & 05:34 & 1 &
900 & N \\ & & & 17/01/15 & 00:20 & 01:40 & 8 & 3840 & N \\ V482 Cam &
NL SW & 3.21 & 18/01/15 & 02:49 & 04:06 & 8 & 3840 & N \\ V425 Cas &
NL VY & 3.59 & 04/08/14 & 02:15 & 03:14 & 6 & 2880 & N \\ & & &
27/10/15 & 22:43 & 23:29 & 3 & 2700 & N \\ CH Crb & NL UX SW & 3.49 &
23/02/15 & 06:09 & 06:24 & 1 & 900 & N \\ V1500 Cyg & Na NL AM AS & 3.35
& 01/08/14 & 02:23 & 02:57 & 3 & 2100 & N \\ V2275 Cyg & Na IP & 7.55
& 02/08/14 & 03:28 & 04:07 & 2 & 1800 & N \\ & & & 03/08/14 & 23:43 &
00:22 & 4 & 1920 & N \\ CM Del & NL UX VY & 3.89 & 02/08/14 & 01:12 &
02:37 & 8 & 3840 & N \\ MN Dra & DN SU ER & 2.40 & 02/08/14 & 23:20 &
01:03 & 8 & 3840 & N \\ OZ Dra & NL UX SW & 3.28 & 18/01/15 & 05:16 &
06:17 & 6 & 2880 & N \\ & & & 02/03/15 & 02:25 & 03:27 & 4 & 3600 & N
\\ V1084 Her & NL SW NS & 2.89 & 02/03/15 & 03:31 & 04:34 & 4 & 3600 &
N \\ BH Lyn & NL SW SH NS & 3.74 & 16/01/15 & 01:11 & 02:39 & 8 & 3840
& N \\ & & & 29/01/15 & 01:27 & 01:42 & 1 & 900 & N \\ BP Lyn & NL UX
SW & 3.67 & 17/01/15 & 02:20 & 03:39 & 8 & 3840 & N \\ & & & 29/01/15
& 01:45 & 02:00 & 8 & 3840 & N \\ HQ Mon & NL UX & 7.58 & 16/12/14 &
04:15 & 04:30 & 1 & 900 & N \\ V380 Oph & NL VY SW NS & 3.70 & 28/05/15
& 23:58 & 01:00 & 4 & 3600 & N \\ V1193 Ori & NL UX SW NS & 3.96 &
17/01/15 & 21:32 & 23:10 & 8 & 3840 & N \\ LQ Peg & NL VY SH NS & 2.99
& 02/08/14 & 02:44 & 03:22 & 4 & 1920 & N \\ FY Per & NL VY & 6.20 &
15/01/15 & 22:05 & 23:36 & 8 & 3840 & N \\ LX Ser & NL VY SW & 3.80 &
03/08/14 & 21:12 & 22:33 & 8 & 3840 & N \\ & & & 17/01/15 & 06:03 &
06:55 & 3 & 2700 & N \\ & & & 18/01/15 & 06:19 & 06:24 & 3 & 180 & N
\\ & & & 02/03/15 & 04:38 & 05:40 & 4 & 3600 & N \\ RW Sex & NL UX &
5.88 & 17/01/15 & 03:44 & 05:05 & 8 & 3840 & N \\ & & & 29/01/15 &
03:18 & 03:33 & 8 & 3840 & N \\ SW Sex & NL UX SW & 3.24 & 16/01/15 &
04:50 & 06:10 & 8 & 3840 & N \\ V1294 Tau & NL VY SW & 3.59 & 16/01/15
& 21:26 & 22:51 & 8 & 3840 & N \\ & & & 13/10/15 & 02:07 & 02:58 & 3 &
2700 & N \\ RW Tri & NL UX SW & 5.57 & 04/08/14 & 03:46 & 04:46 & 6 &
2880 & N \\ DW UMa & NL SW SH NS & 3.28 & 16/01/15 & 03:28 & 04:46 & 8
& 3840 & N \\ & & & 29/01/15 & 03:37 & 03:52 & 1 & 900 & N \\ & & &
23/02/15 & 04:12 & 04:59 & 3 & 2700 & N \\ LN UMa & NL VY SW & 3.47 &
18/01/15 & 04:11 & 05:08 & 6 & 2880 & N \\ UX UMa & NL UX & 4.72 &
16/01/15 & 06:15 & 06:59 & 6 & 2880 & N \\ & & & 17/01/15 & 05:09 &
05:57 & 6 & 2880 & N \\ & & & 01/03/15 & 03:15 & 04:16 & 6 & 3600 & N
\\ SS UMi & DN SU ER & 1.63 & 01/08/14 & 21:32 & 22:35 & 4 & 3600 & N
\\ & & & 02/08/14 & 21:16 & 21:29 & 1 & 240 & N \\ HS0220+0603 & NL UX
SW & 3.58 & 17/01/15 & 19:55 & 21:19 & 8 & 3840 & N \\ HS0229+8016 &
NL UX VY & 3.88 & 16/01/15 & 19:59 & 21:22 & 8 & 3840 & N
\\ HS0455+8315 & NL VY SW & 3.57 & 18/01/15 & 00:03 & 01:23 & 8 & 3840
& N \\ HS1813+6122 & NL UX SW NS & 3.55 & 13/10/15 & 21:37 & 22:24 & 3
                            & 2700 & N \\
 & & & 27/10/15 & 21:14 &  21:29 & 1 & 900 & N \\  
\hline
  \label{tab:journal}
\end{tabular}
\end{center}
\end{minipage}
\end{table*}

\begin{table*}
  \contcaption{}
   \begin{minipage}{160mm}
     \begin{center}
\begin{tabular}{llccccccc}
  \hline \hline & & & & & & & & \\ 
 \multicolumn{1}{l}{Object} &
\multicolumn{1}{l}{Classification} & \multicolumn{1}{c}{Orbital} &
\multicolumn{1}{c}{Date} & \multicolumn{1}{c}{UTC} &
\multicolumn{1}{c}{UTC} & \multicolumn{1}{c}{Number of} &
\multicolumn{1}{c}{Total exposure} & \multicolumn{1}{c}{Visible} \\ &
& \multicolumn{1}{c}{period (hrs)} & & \multicolumn{1}{c}{start} &
\multicolumn{1}{c}{end} & \multicolumn{1}{c}{exposures} &
 \multicolumn{1}{c}{time (secs)} & \multicolumn{1}{c}{shell?} \\
 & & & & & & & & \\ \hline
 J0506+7725 & DN SU & 1.62 & 28/10/15 & 03:11 & 04:13 & 4 & 3600 & N
\\ J0809+3814 & NL SW & 3.21 & 29/01/15 & 01:07 & 01:22 & 1 & 900 & N
\\ J0928+5004 & NL UX & 10.04 & 29/01/15 & 02:04 & 02:19 & 1 & 900 & N
\\ J1429+4145 & NL & 1.64 & 23/02/15 & 05:03 & 06:05 & 4 & 3600 & N
\\ J1924+4459 & NL SW SH NS & 2.75 & 27/10/15 & 21:36 & 22:38 & 4 &
3600 & N \\ Leo5 & DN ZC & 3.51 & 28/05/15 & 22:44 & 23:46 & 4 & 3600
& N \\ LSIV-083 & NL UX & 4.69 & 29/05/15 & 01:11 & 02:13 & 4 & 3600 &
N \\ RXJ0524+4244 & NL AM AS & 2.62 & 16/01/15 & 23:17 & 00:17 & 8 & 3840 & N \\ \hline
\hline
\end{tabular}
\end{center}
\end{minipage}
\end{table*}

\subsection{Notes on Individual Targets}

There are some noteworthy features in the images and radial profiles of
individual targets that we discuss below:

\begin{itemize}
\item BZ Cam has a radial profile above the field stars because of emission from
  the pre-existing bow-shock nebula. \citet{hoffmann20} proposed that this
  system could be a recurrent nova instead of a NL.
\item The asynchronous polar V1500 Cyg has a pronounced peak in its radial
  profile at around 5$\arcsec$, from the nova event in 1975 \citep{lindegren75}.
\item V2275 Cyg was discussed extensively in S15, where we discovered a number
  of H$\alpha$ blobs in IPHAS images that appeared to move over time. We
  attributed this apparent motion to a light echo from the 2001 nova eruption.
  The blobs are not discernible in our image, which was taken six years after
  the last IPHAS image used in S15.
 \item CM Del lies in a crowded field, and is blended with a nearby star. Hence
   its radial profile is artificially enlarged.
\item FY Per has a number of nearby stars and hence the radial profile has an
  unreliable shape.  
\item J1429+4145 is extremely faint in our image and we were unable to derive a
  radial profile.  
\end{itemize}

A number of the brighter targets were over--exposed, which leads to the
saturation of the central pixels and gives the radial profile a flat--top
profile. The targets affected were HL Aqr, UU Aqr, V1084 Her, RW Tri and
LSIV-083. Despite being saturated, the wings of the radial profiles are still
useful for detecting small shells. None of the other radial profiles of
our targets showed any significant deviation from the field stars, implying that
there were no small shells around any of the targets.

\subsection{Combined results of our two campaigns and other recent work}

In Table \ref{tab:tottab} we summarise the combined results from S15 and our
second campaign reported in this paper, which shows that we have examined a
total of 132 CVs, including 51 NLs. Table \ref{tab:tottab} also shows the
results from two other similar campaigns by \citet{schmidtobreick15} and
\citet{pagnotta16}, neither of whom found shells around the CVs they examined.

\begin{table*}
\caption[]{Summary of our search for nova shells.}
\begin{center}
\begin{tabular}{lrrrrrr}
\hline\hline \multicolumn{1}{l}{ } & \multicolumn{1}{r}{Nova-like} &
\multicolumn{1}{r}{Polars \&} & \multicolumn{1}{r}{Asynchronous} &
\multicolumn{1}{r}{Dwarf} & \multicolumn{1}{r}{Old} &
\multicolumn{1}{r}{Total} \\ \multicolumn{1}{l}{ } &
\multicolumn{1}{r}{Variables} & \multicolumn{1}{r}{Intermediate} &
\multicolumn{1}{r}{Polars} & \multicolumn{1}{r}{Novae} &
\multicolumn{1}{r}{Novae} & \multicolumn{1}{r}{}
\\ \multicolumn{1}{l}{ } & \multicolumn{1}{r}{} &
\multicolumn{1}{r}{Polars} & \multicolumn{1}{r}{} &
\multicolumn{1}{r}{} & \multicolumn{1}{r}{} & \multicolumn{1}{r}{}
\\ \hline Original paper \citep{sahman15} & & & & & & \\ WHT Targets &
22 & 1 & 2 & 2 & 4 & 31 \\ IPHAS Targets & 5 & 10 & 2 & 34 & 23 & 74
\\ \multicolumn{1}{l}{less: Duplicated objects}& --3 & & & & --1 & --4
\\ \hline \multicolumn{1}{l}{Total} & 24 & 11 & 4 & 36 & 26 & 101
\\ \hline This Paper & 39 &1 & 3 & 3 & 1 & 47 \\ less: Duplicated
objects & --12 &--1 & --3 & & & --16 \\ \hline
\multicolumn{1}{l}{Total} & 51 & 11 & 4 & 39 & 27 & 132 \\ { } & & & &
& & \\ \citet{schmidtobreick15} & 5 & & & 10 & & 15
\\ \citet{pagnotta16} & & 1 & 3 & &  & 4 \\ less: Duplicated objects
& & & --3 & & & --3 \\ \hline \multicolumn{1}{l}{Grand total} & 56 &
12 & 4 & 49 & 27 & 148 \\ \hline\hline
\end{tabular}
\end{center}
\label{tab:tottab}
\end{table*}

\section{Discussion}
\label{sec:disc}

Our new survey brings the total number of NLs that have been observed to 56 (see
Table \ref{tab:tottab}), with one nova shell discovered around the nova-like
V1315 Aql \citep{sahman18}. What can we deduce from this? Given the much
increased rate of nova detection in the last century, let us assume that any
novae eruptions in these NLs that occurred in the last $\sim$100 years would
have been observed. These would now be classified as old novae in the RK
catalogue and hence would not appear in our sample of NLs. We also know from our
simulations (see Section 2.1) that our observations are not sensitive to shells
older than $\sim$200 yrs. Hence our search for nova shells around NLs is only
likely to find shells between $\sim$100 and $\sim$200 years old. We found one
shell in this $\sim$100-year window, out of 56 NLs surveyed, indicating that the
lifetime of the NL phase lasts approximately $\sim$5,600\,yrs.

If we include the discovery of the nova shell around IPHASX J210204.7+471015 by
\citet{guerrero18}, who used a similar setup, in our calculation then we have
two NLs with shells from a total of 57 NLs. Assuming the $\sim$100-year
visibility window as discussed in the previous paragraph, this would imply a NL
lifetime of $\sim$3,000 yrs. This result is broadly consistent with the
order-of-magnitude estimate of 1,000 years derived by \citet{patterson13} for
the NL phase of CVs.

\citet{schmidtobreick15} found no shells around the 15 CVs they examined. They
derived a lower limit of 13,000 years for the overall nova recurrence time. This
is consistent with the lifetime of the NL phase of $\sim$3,000 yrs that we have
derived, as one would expect that the NL phase should be shorter than the overall
nova recurrence time, assuming that the cyclical evolution theory is correct.

The latest models by \citet{hillman20} show that the behaviour of a CV during a
nova cycle is largely determined by the masses of the two component stars. They
modelled four CVs, with WDs of mass 0.7 M$_{\sun}$ and 1.0 M$_{\sun}$ and
secondaries of mass 0.45 M$_{\sun}$ and 0.7 M$_{\sun}$. They found that the high
mass pair were most likely to exhibit high mass transfer rates, and hence appear
as NLs, whereas the low mass pair only achieved NL mass transfer rates just
prior to a nova eruption, and only during the phase when both magnetic braking
and gravitational wave radiation were operating. This suggests that both V1315
Aql and IPHASX J210204.7+471015 are more likely to harbour high mass components.

The search by \citet{pagnotta16} for nova shells around three asynchronous
polars and one intermediate polar found no shells. Their aim was to test the
theory that the asynchronicity of the WD spin was caused by past nova eruptions
\citep{campbell99}. Our search included three asynchronous polars, V1432 Aql, BY
Cam, and V1500 Cyg. Two of these, V1432 Aql and BY Cam, were observed by
\citet{pagnotta16}. Our observations were not as deep as theirs, and we had a
smaller field of view so, unsurprisingly, we also found no shells.

\section{Conclusions}

We performed a second H$\alpha$-imaging survey to search for old nova shells
around CVs. We imaged 47 CVs with the INT and found no new shells. Assuming that
the nova-induced cycle theory is correct, our results, when combined with our
previous campaign and other recent similar surveys, imply that the nova-like
phase for CVs lasts $\sim$3{,}000 years.

\section*{Acknowledgements}

VSD is supported by the Science and Technology Facilities Council (STFC). The
INT and its service programme are operated on the island of La Palma by the
Isaac Newton Group in the Spanish Observatorio del Roque de los Muchachos of the
Instituto de Astrof\'{i}sica de Canarias. 

\section{Data Availability}

Data available on request.

\newpage

\bibliographystyle{mn2e} \bibliography{abbrev.bib,refs.bib}

\clearpage

\appendix

\section[]{INT Images}
\label{app1}
\begin{figure}
  \centering
  \vspace{10pt}
  \includegraphics[width=80mm,angle=0]{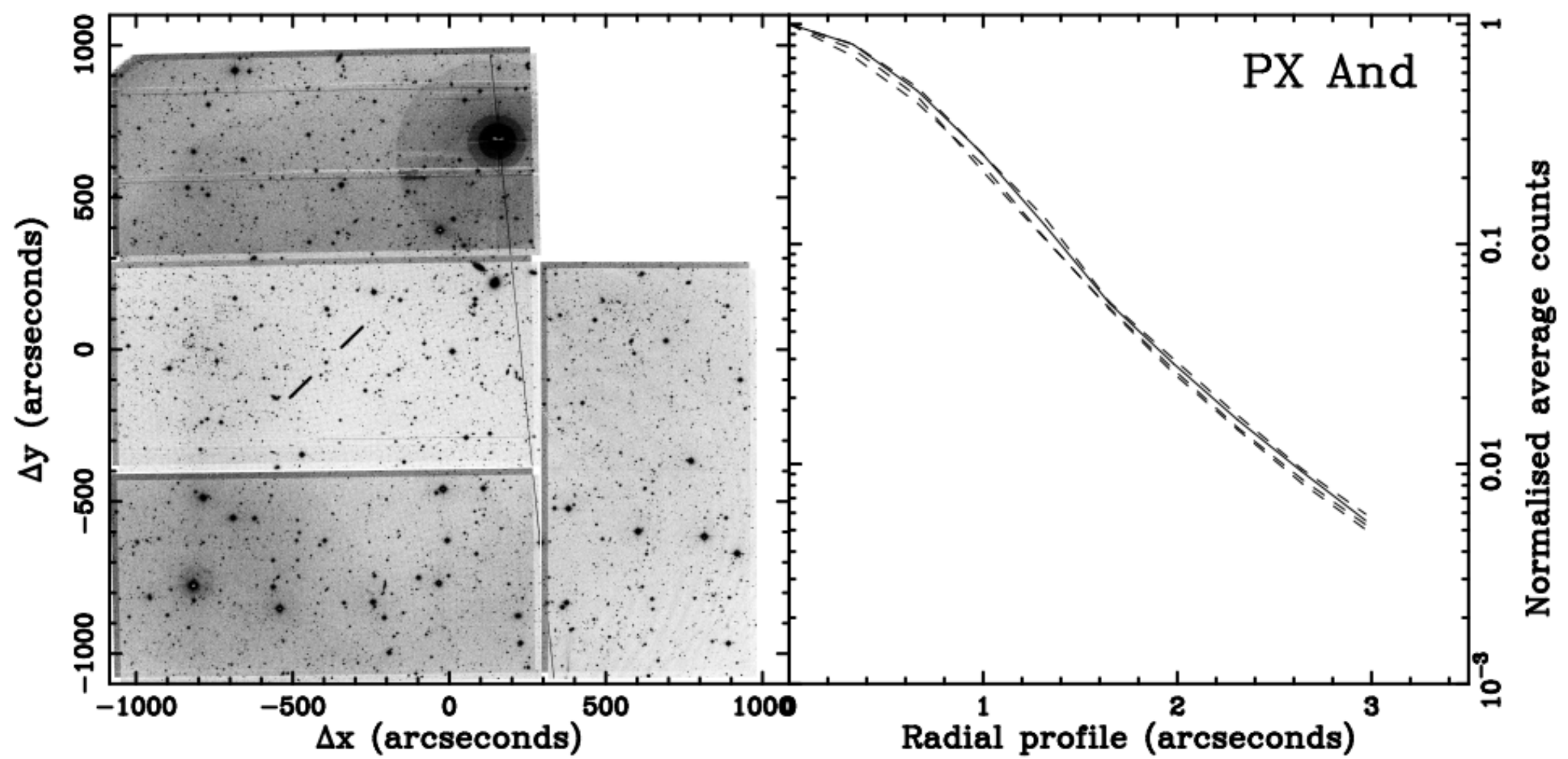}
  \vspace{10pt}
  \includegraphics[width=80mm,angle=0]{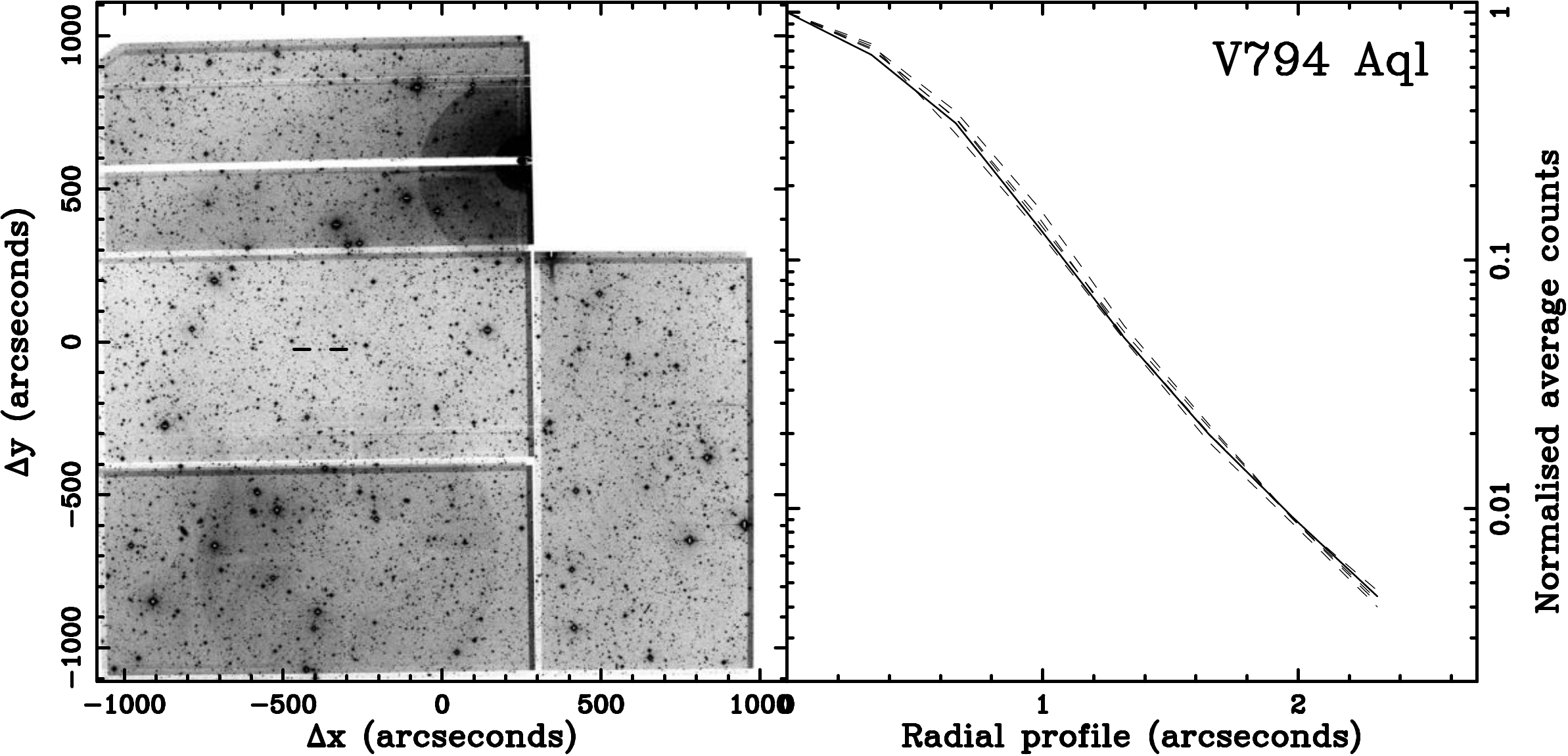} 
  \vspace{10pt}
  \includegraphics[width=80mm,angle=0]{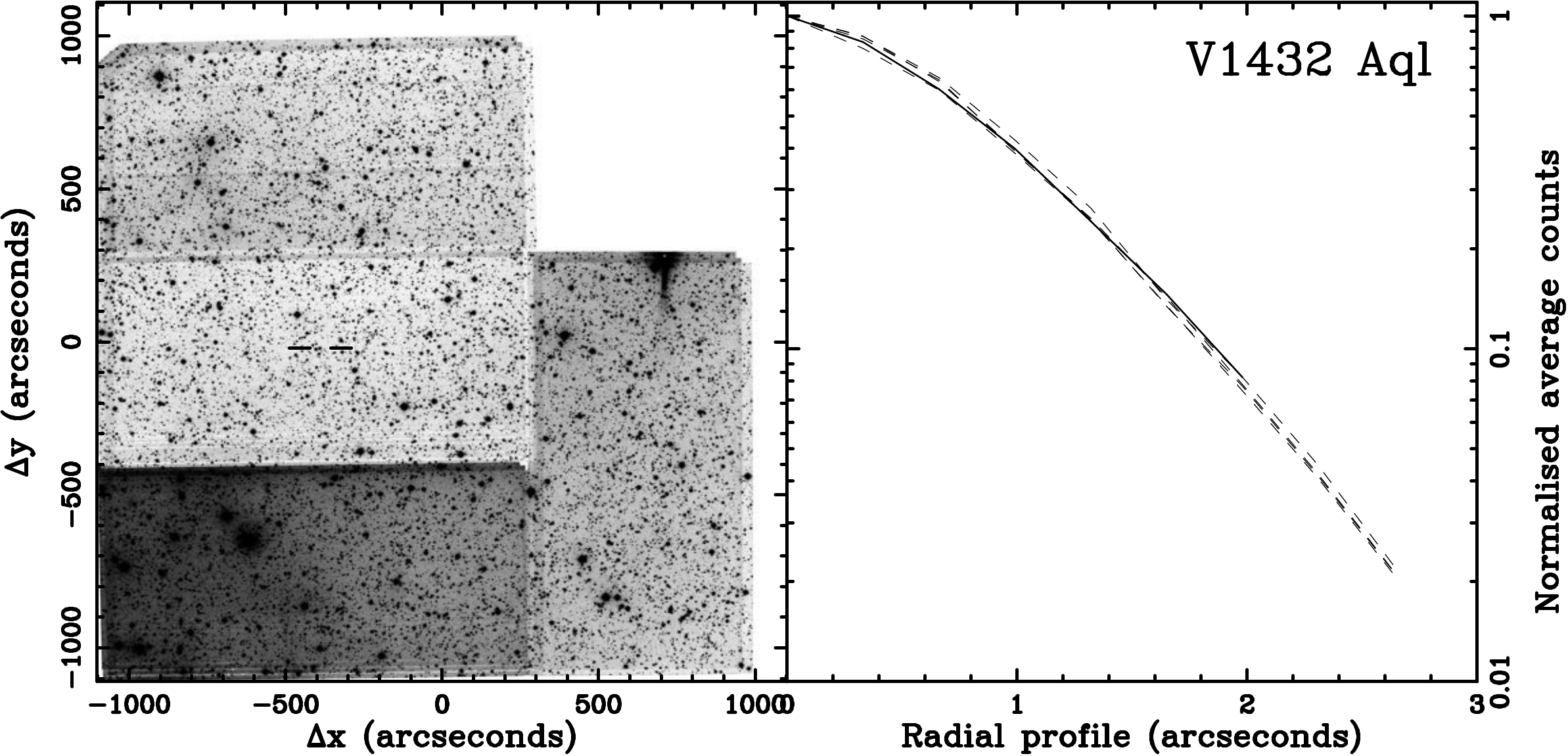}
  \vspace{10pt}
  \includegraphics[width=80mm,angle=0]{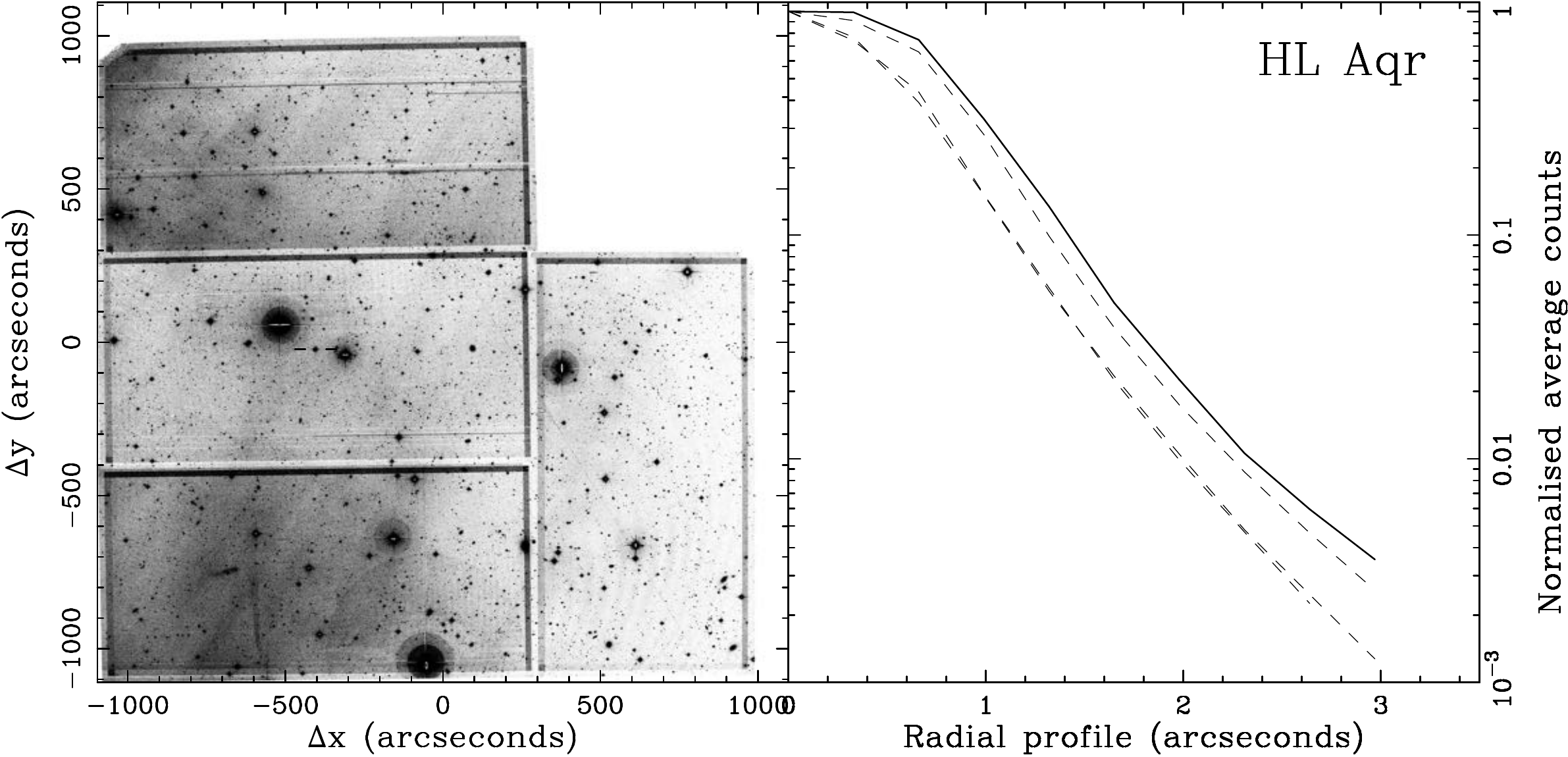}
  \vspace{10pt}
  \includegraphics[width=80mm,angle=0]{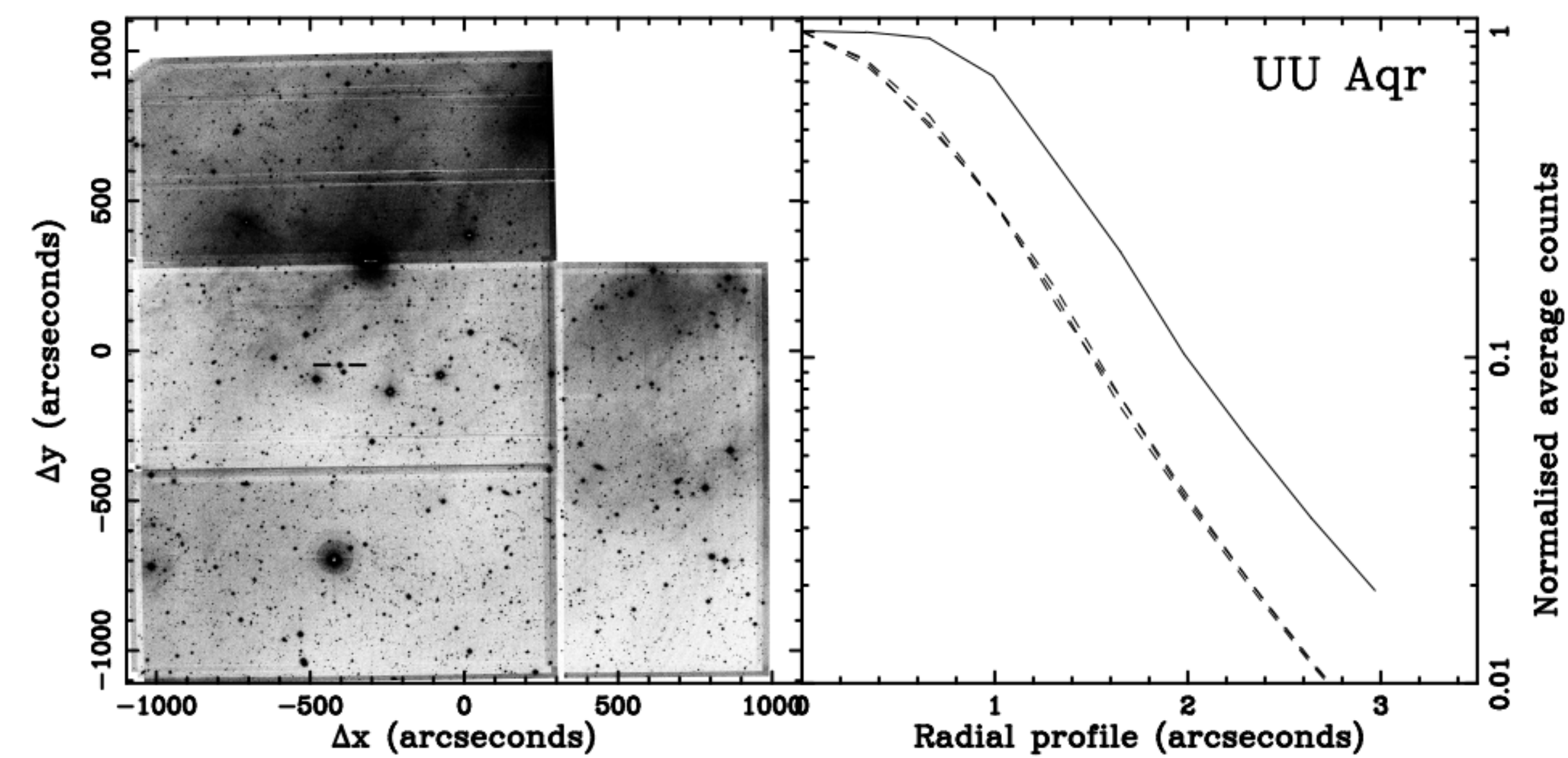} 
   \captionof{figure}{INT images of our target CVs in order of constellation. The CV lies at the centre of the middle CCD on the left-hand side and is indicated by tick marks. In all images, North is up and East is left. Right: Radial profiles of our targets (solid lines) and field stars (dashed lines).}
  \label{fig1}
\end{figure}

\begin{figure}
  \centering
  \includegraphics[width=80mm,angle=0]{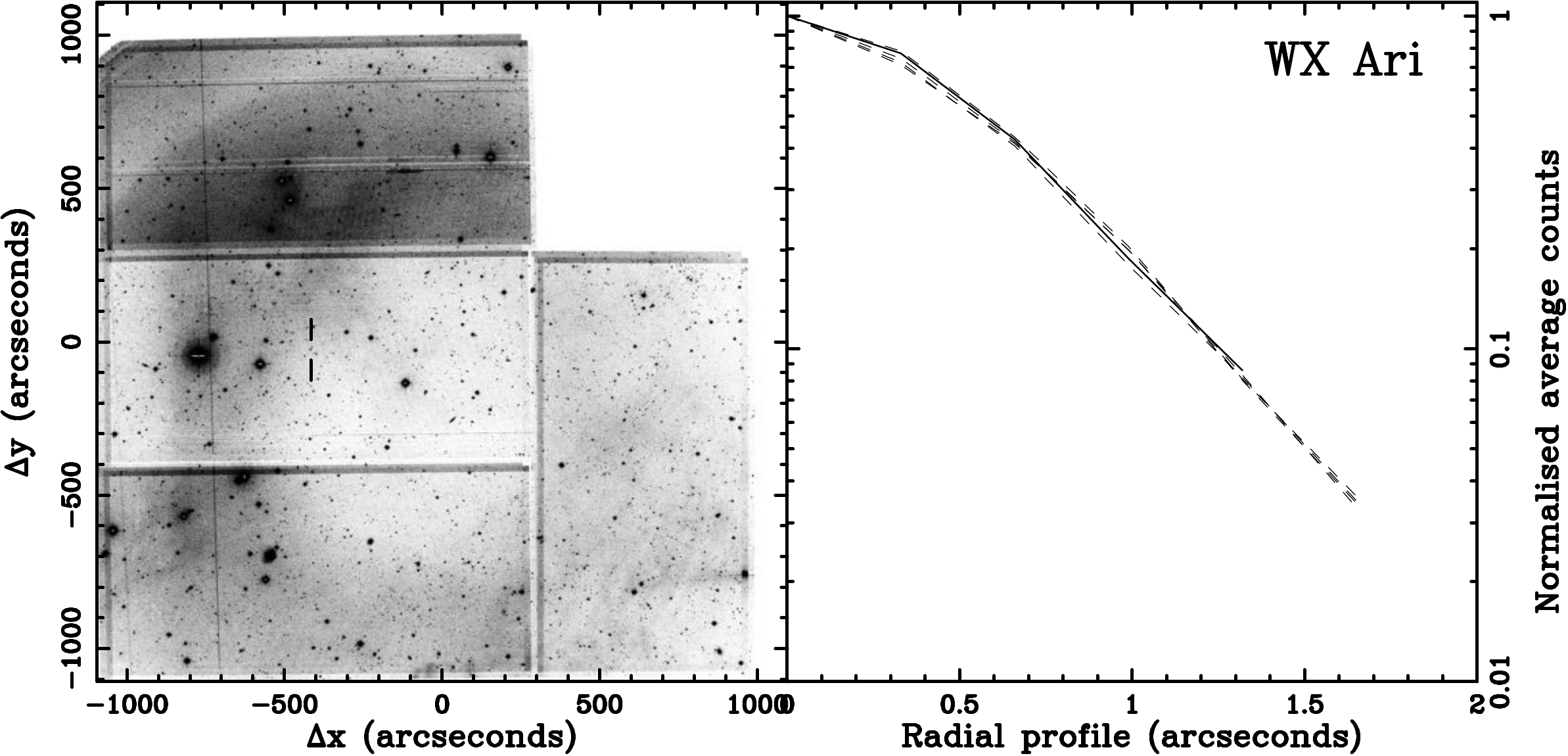}
  \vspace{10pt}
  \includegraphics[width=80mm,angle=0]{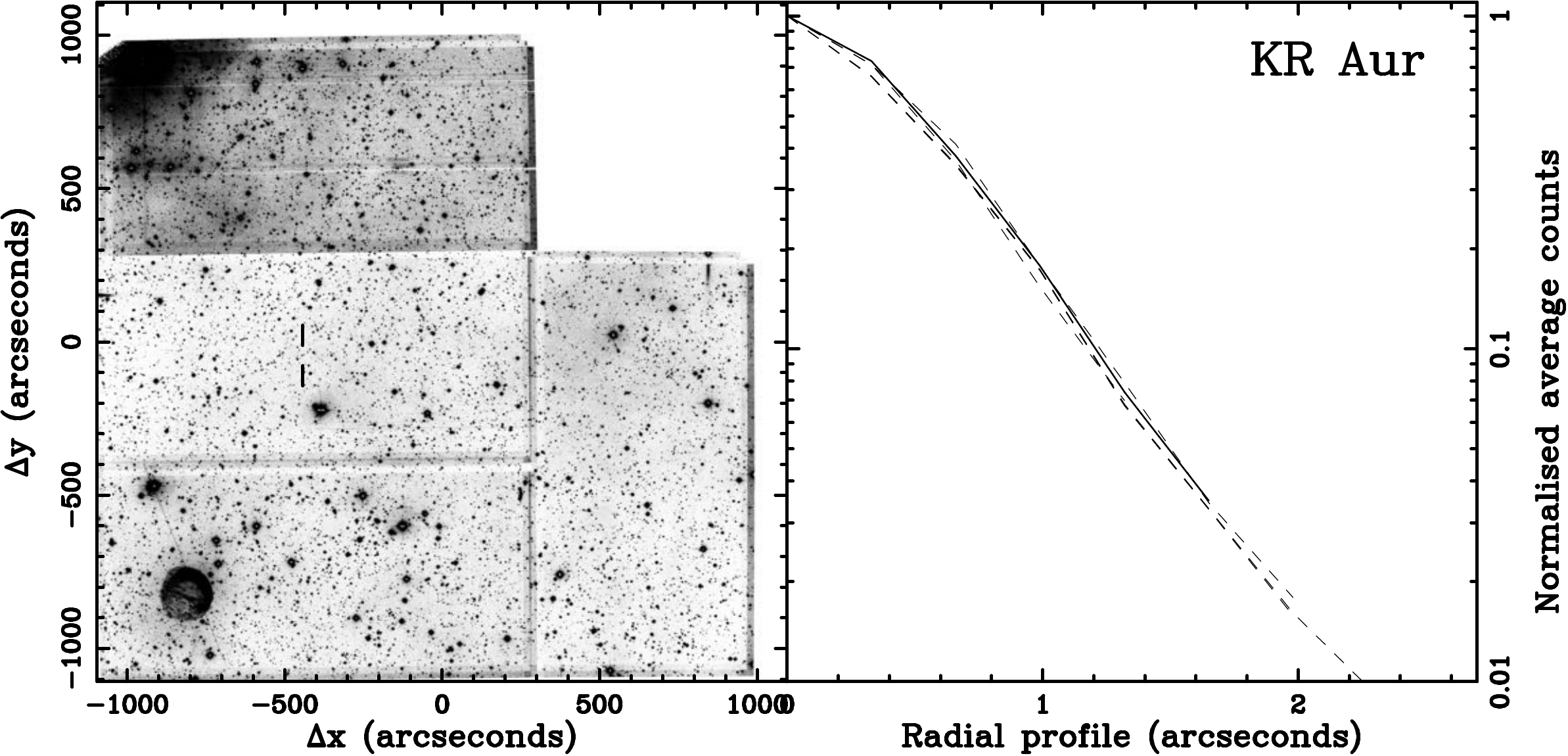}
  \vspace{10pt}
  \includegraphics[width=80mm,angle=0]{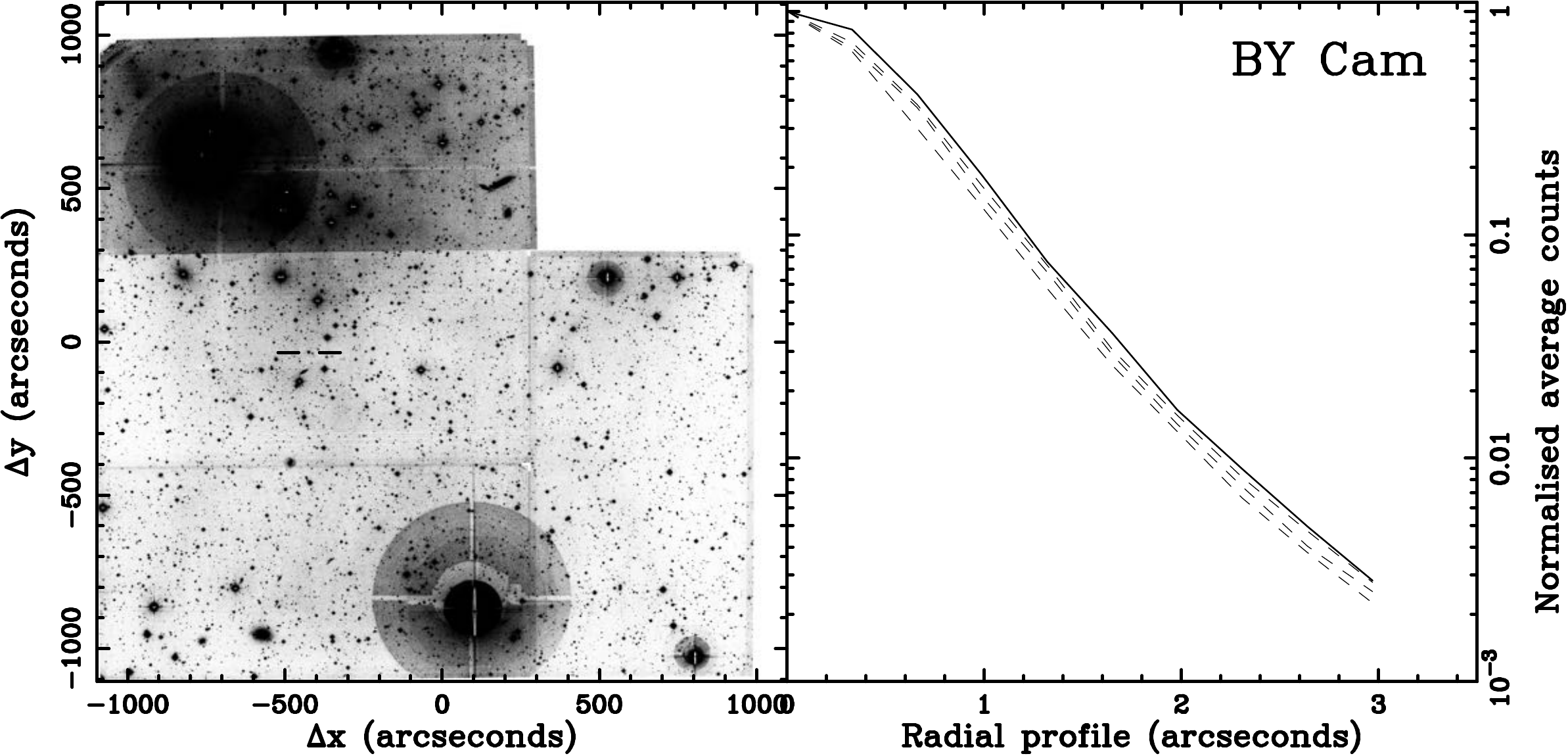}
 \vspace{10pt}
  \includegraphics[width=80mm,angle=0]{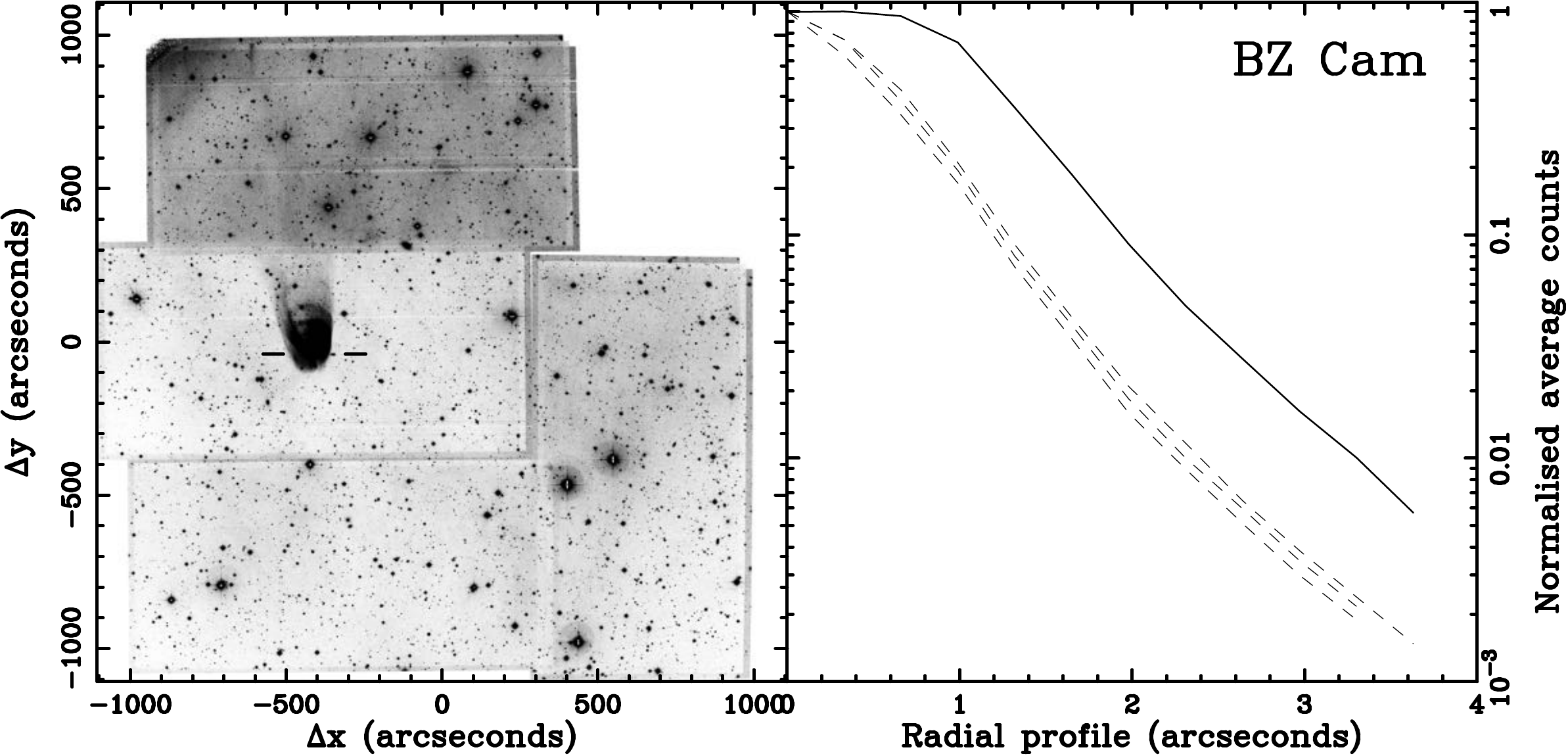}
  \vspace{10pt}
  \includegraphics[width=80mm,angle=0]{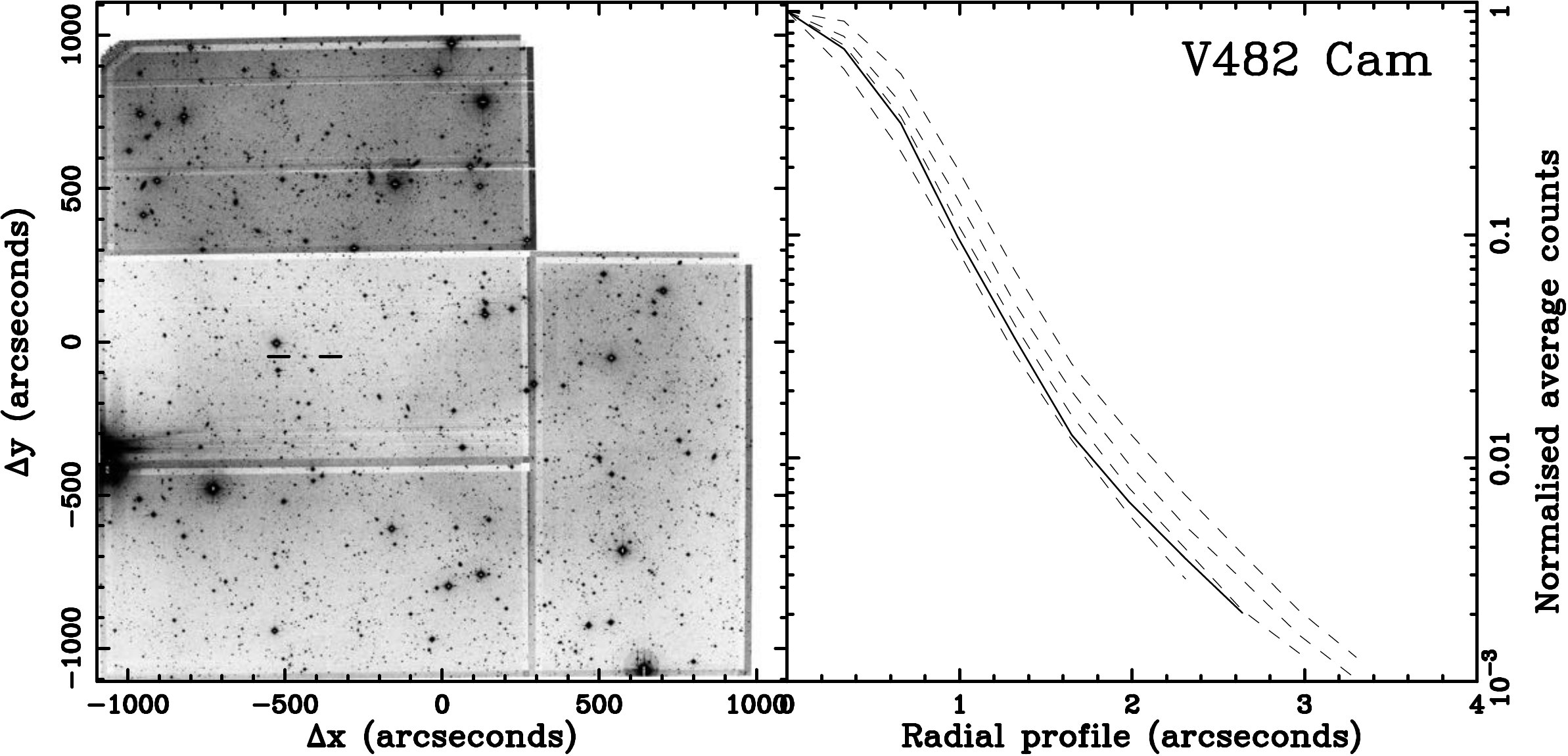}
  \caption{See caption to Figure \ref{fig1} for details.}
\label{fig2}
\end{figure}

\begin{figure}
\centering
  \includegraphics[width=80mm,angle=0]{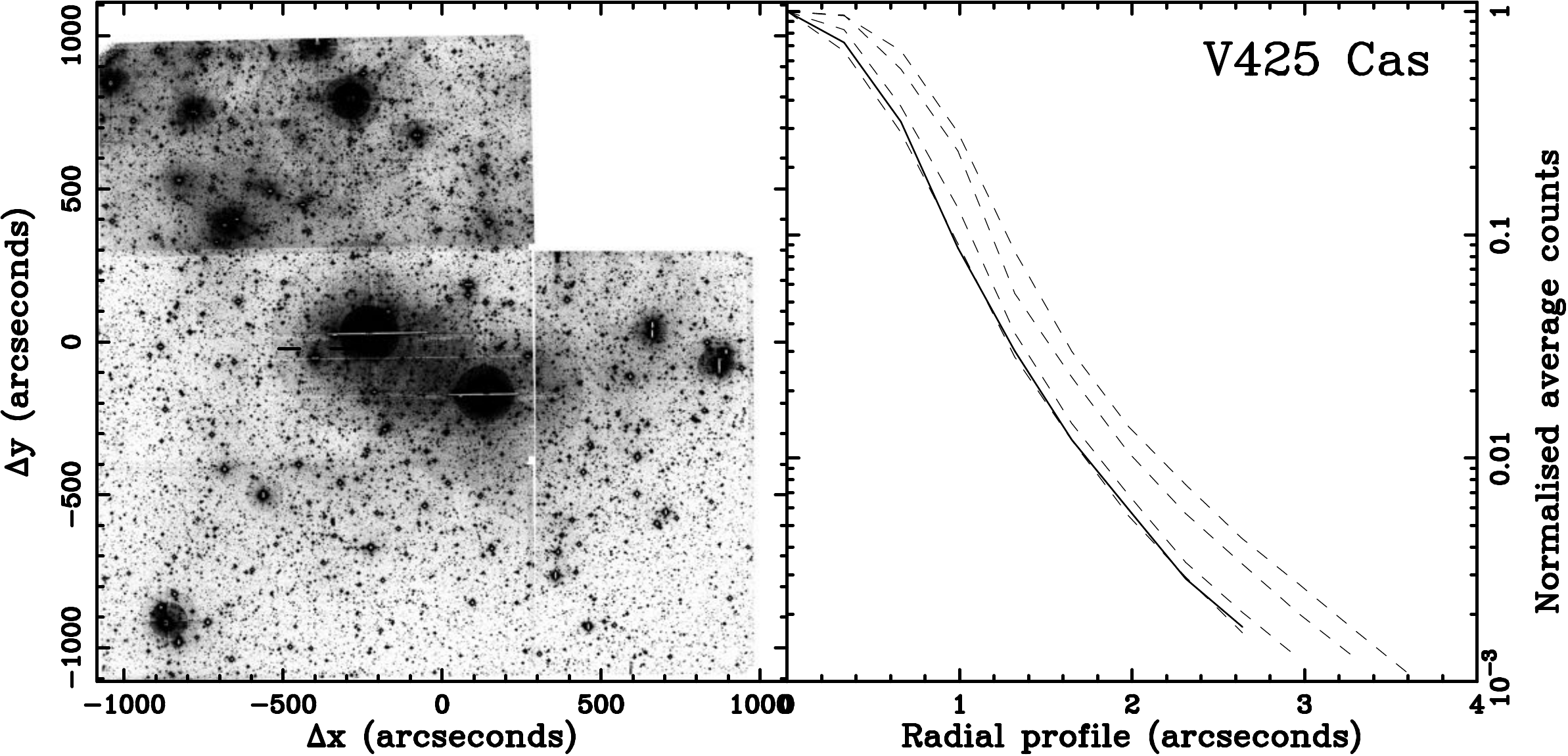}
  \vspace{10pt}
  \includegraphics[width=80mm,angle=0]{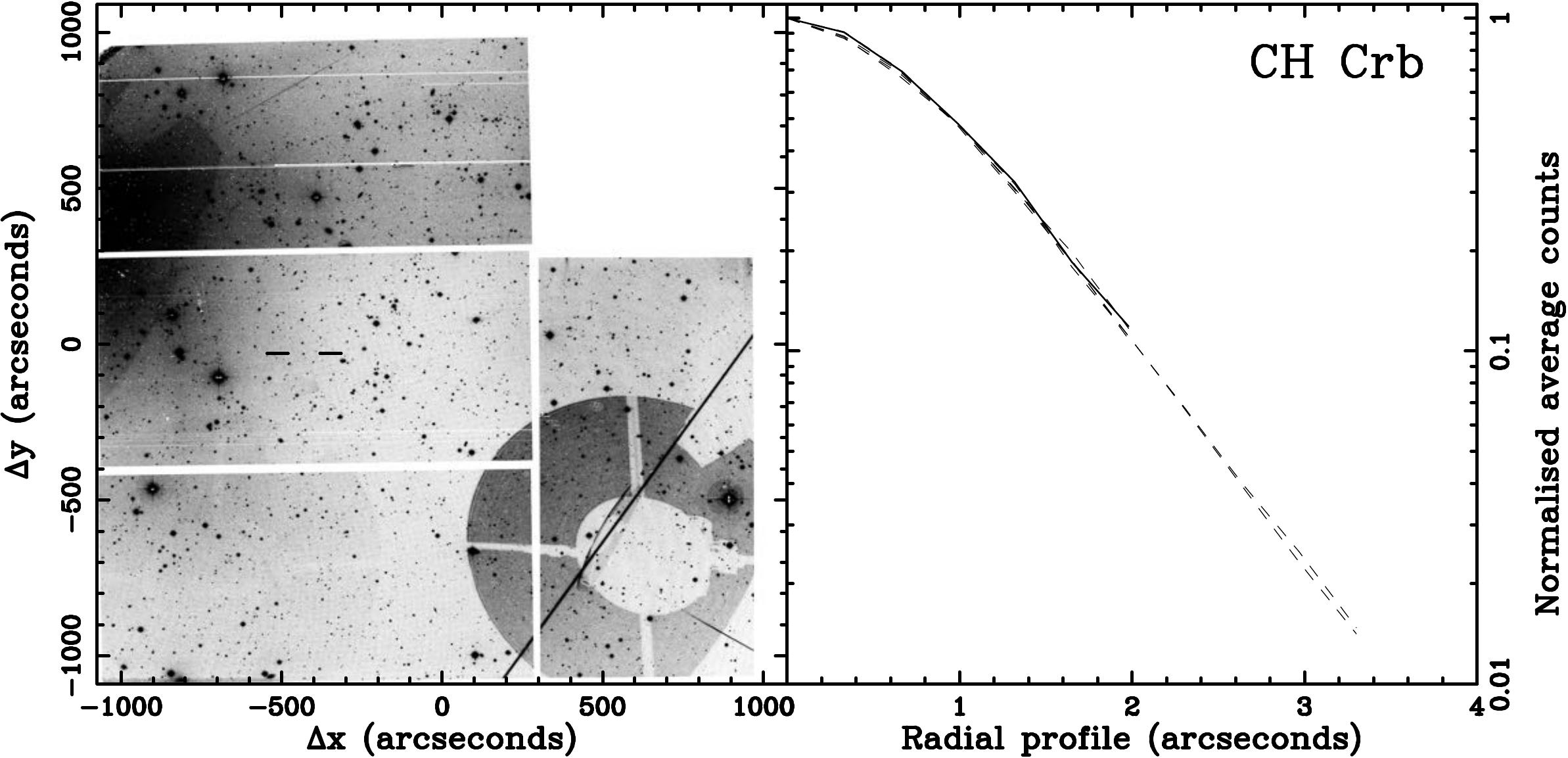}
  \vspace{10pt}
  \includegraphics[width=80mm,angle=0]{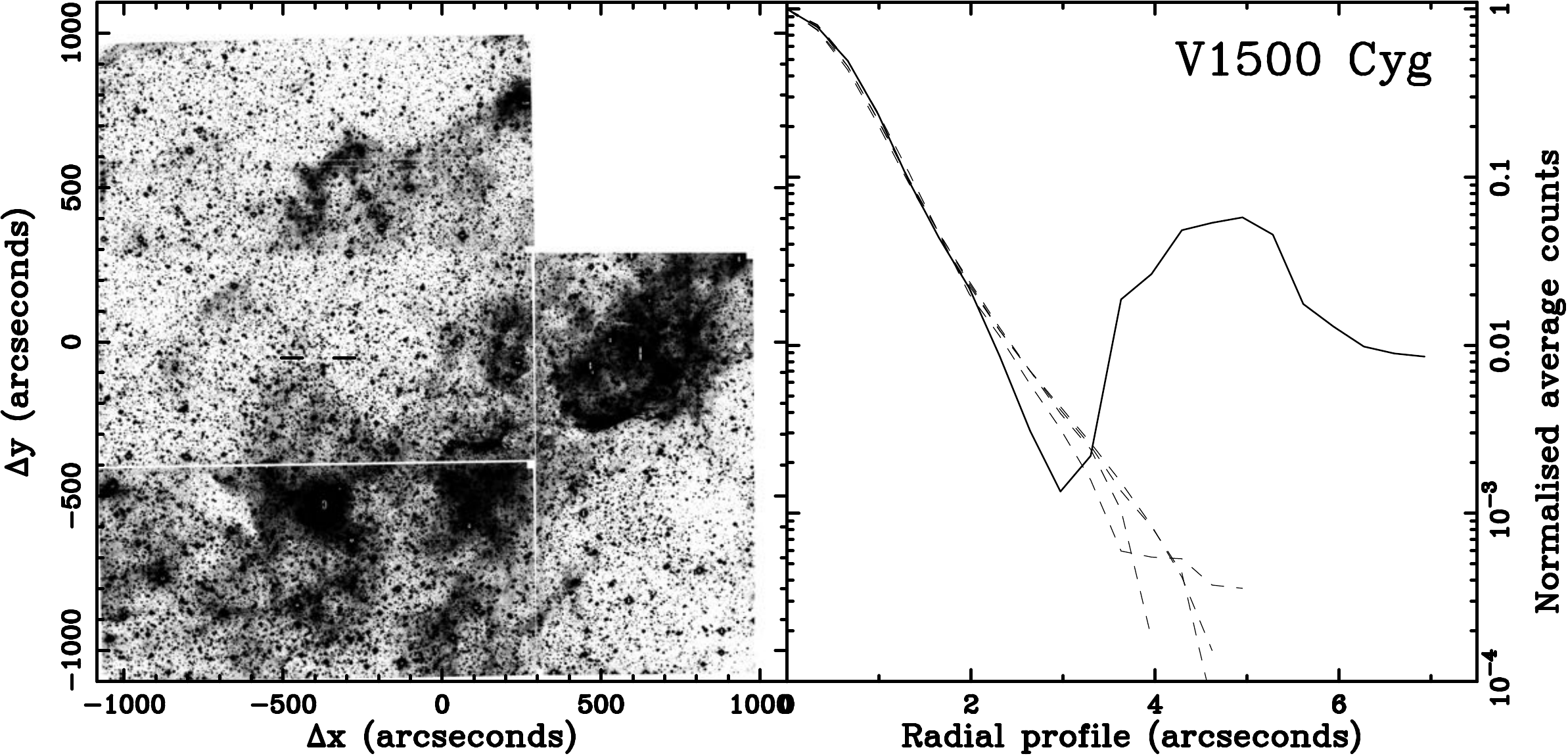}
  \vspace{10pt}
  \includegraphics[width=80mm,angle=0]{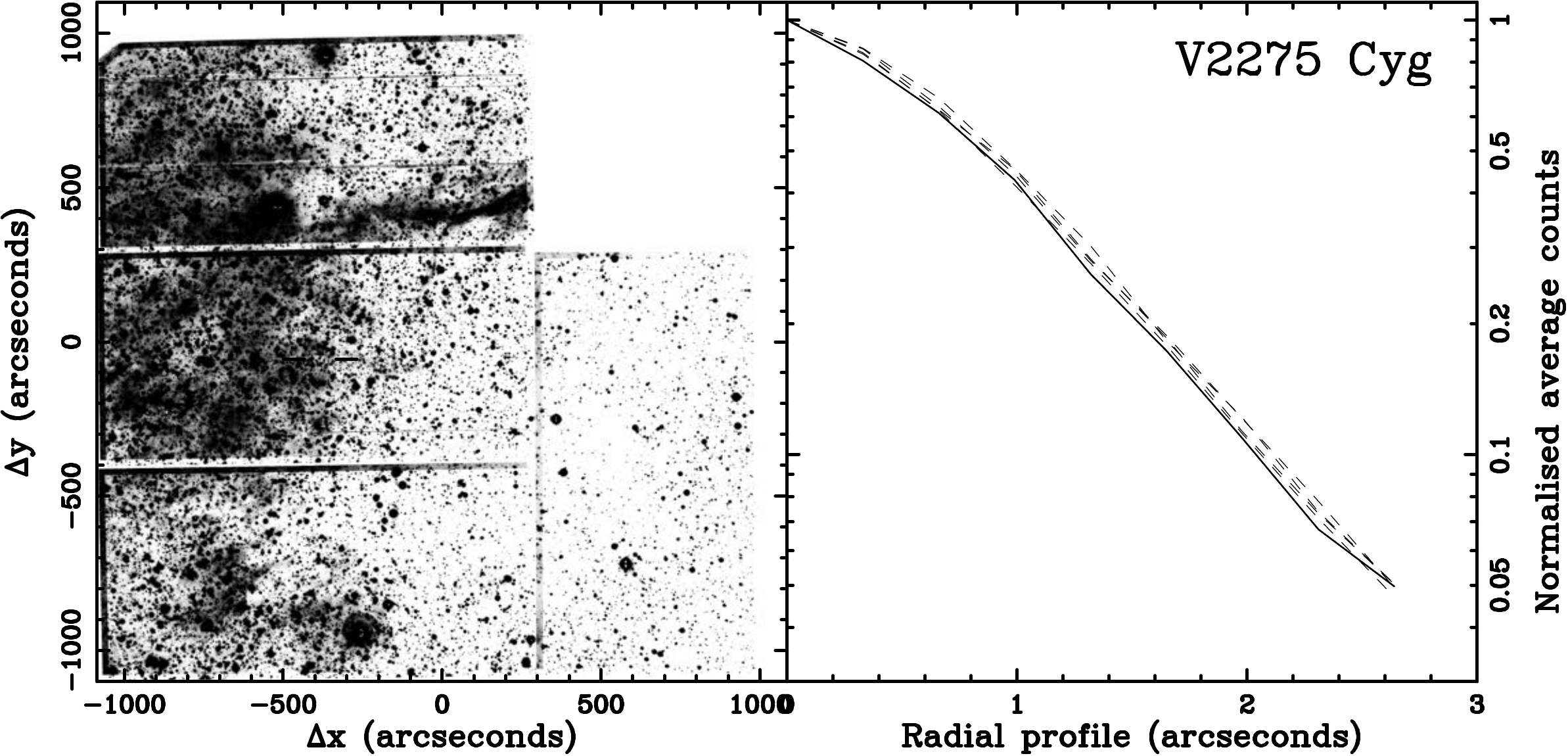}
  \vspace{10pt}
  \includegraphics[width=80mm,angle=0]{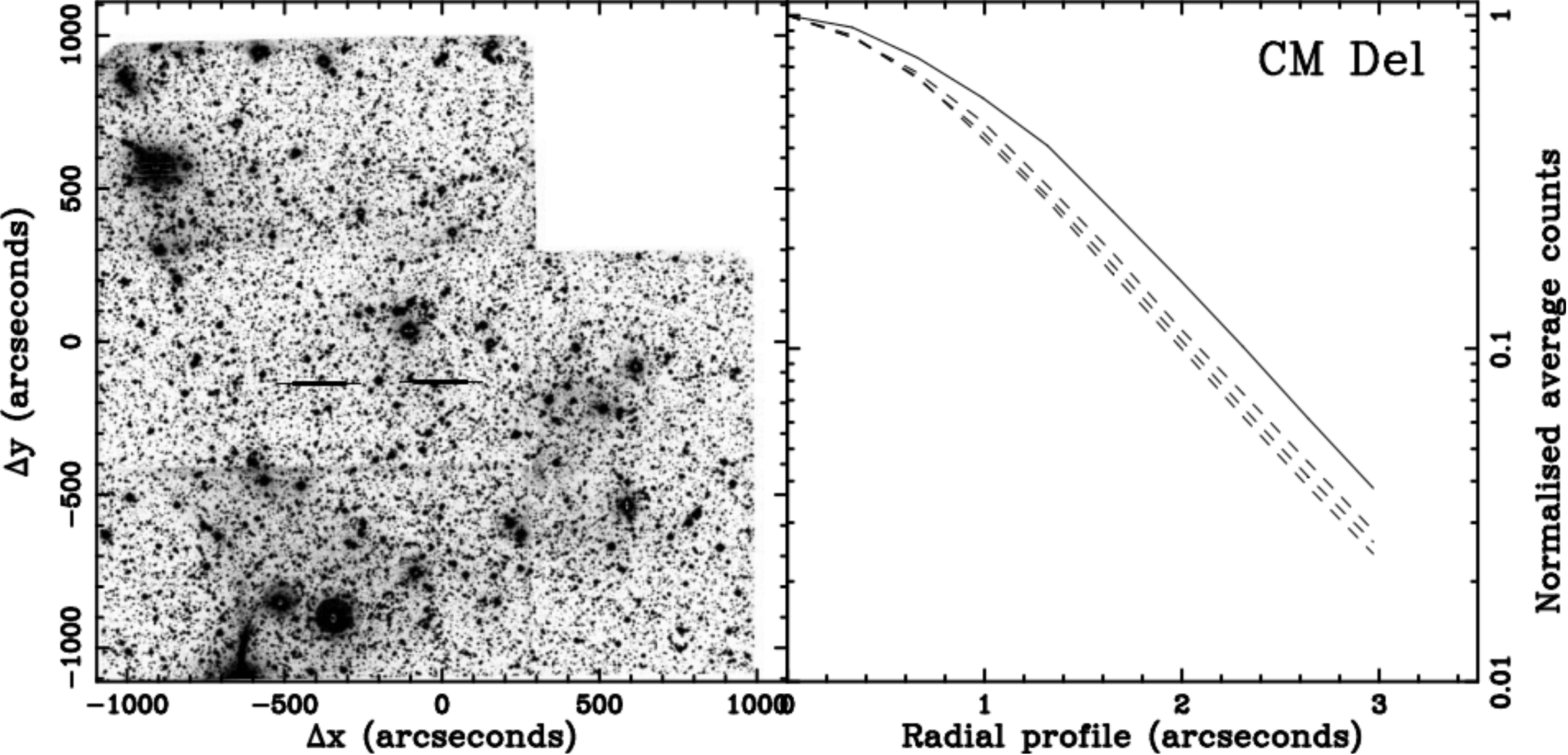}
\caption{See caption to Figure \ref{fig1} for details.}
 \label{fig3}
\end{figure}

\begin{figure}
\centering
  \includegraphics[width=80mm,angle=0]{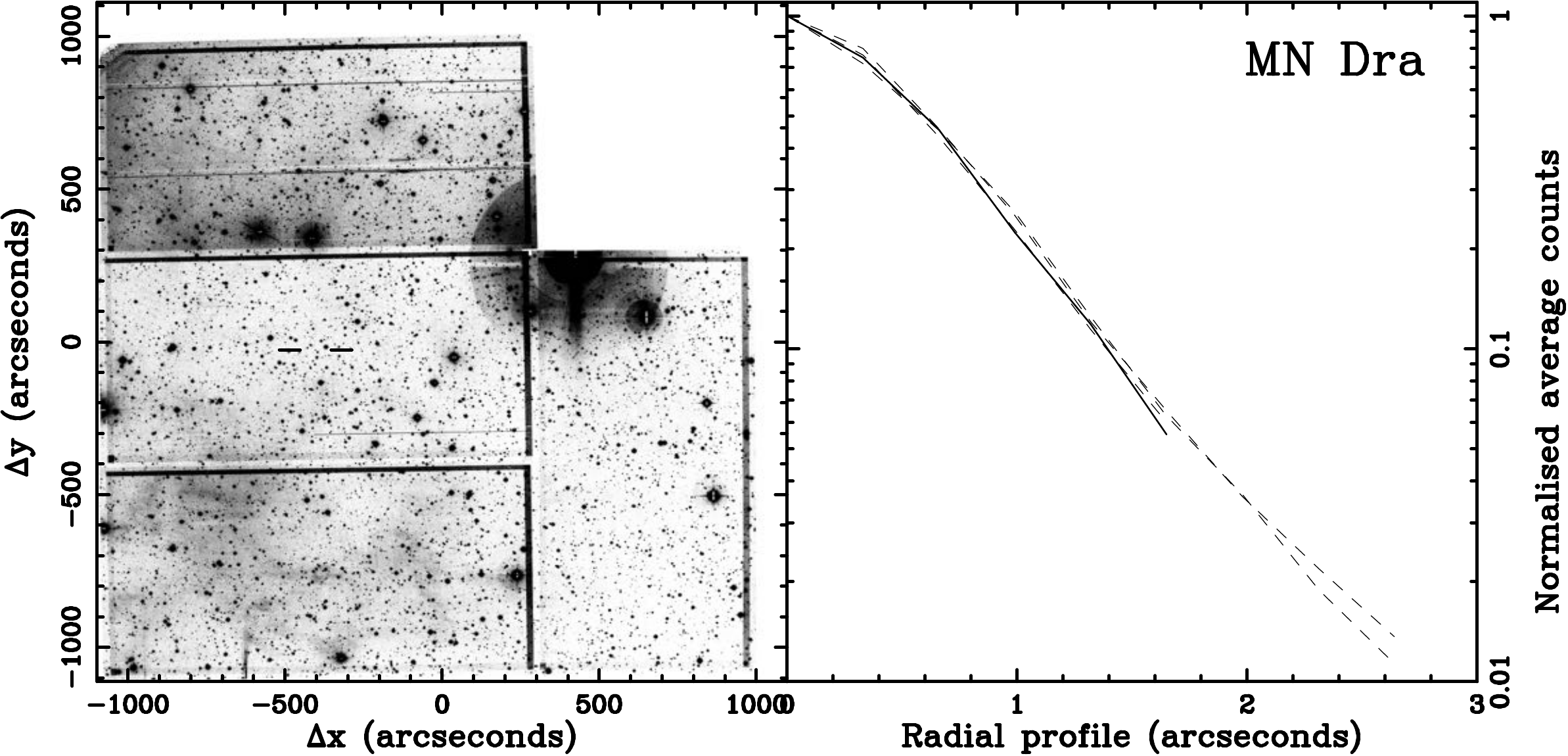}
  \vspace{10pt}
  \includegraphics[width=80mm,angle=0]{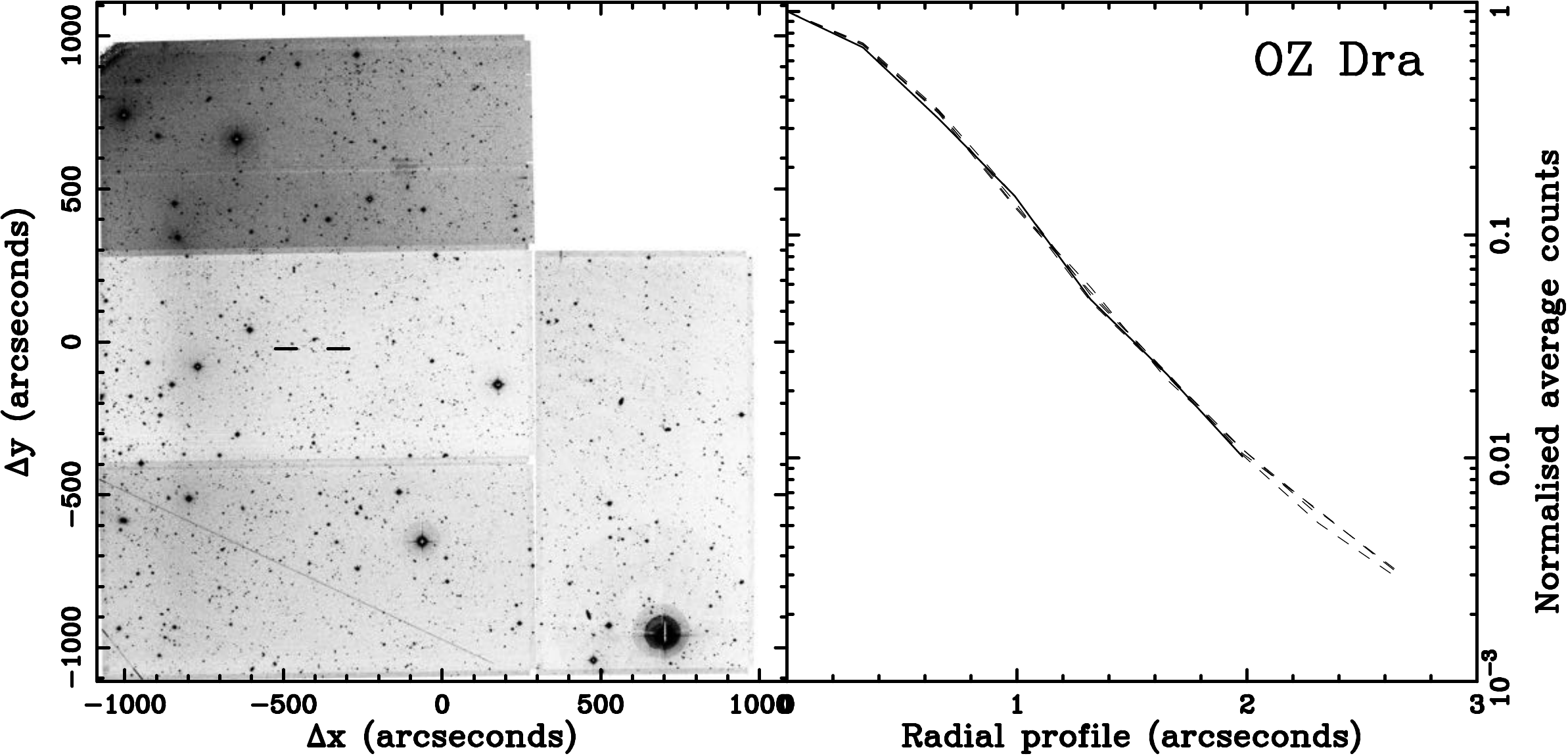}
  \vspace{10pt}
  \includegraphics[width=80mm,angle=0]{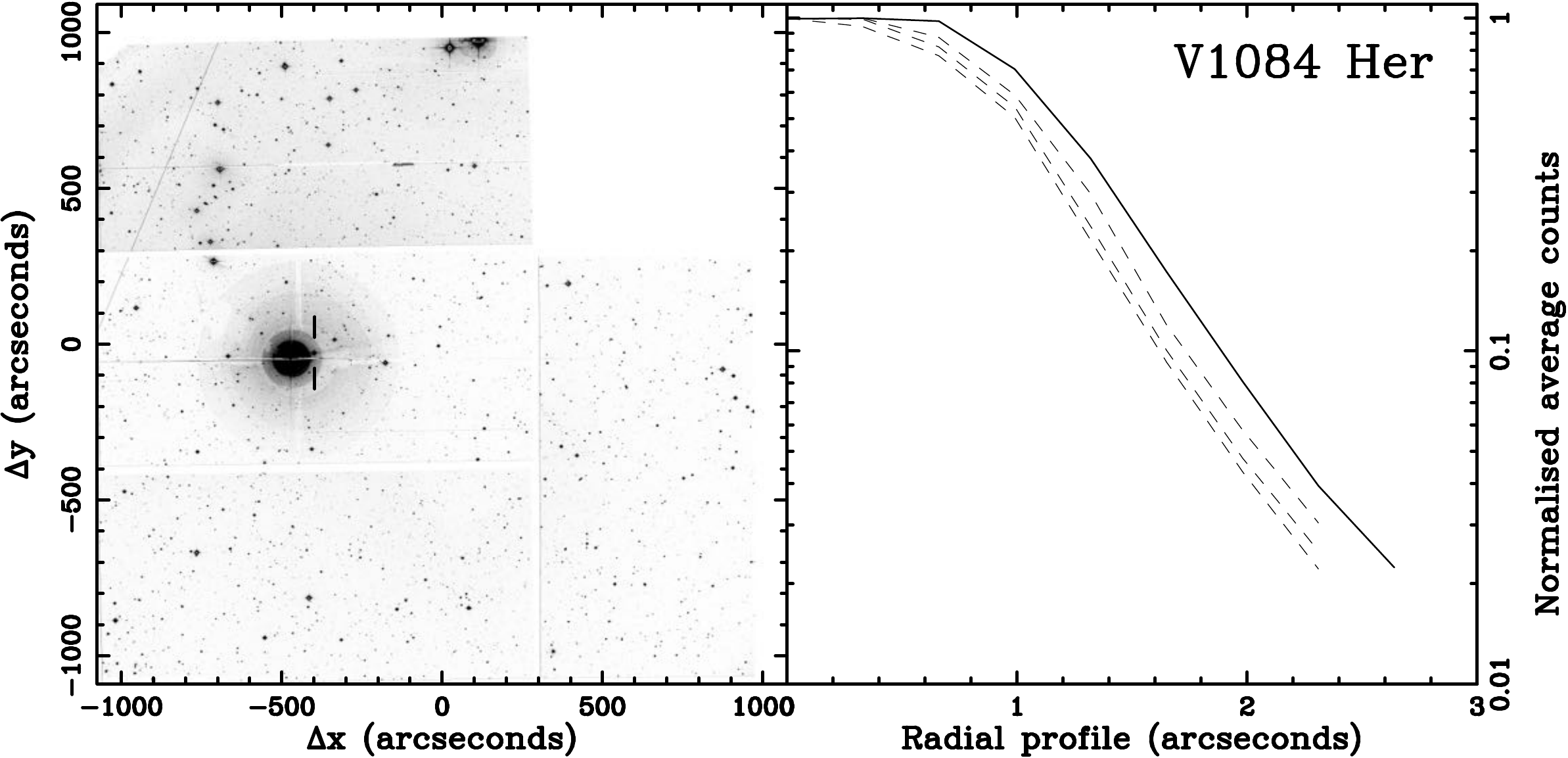}
  \vspace{10pt}
  \includegraphics[width=80mm,angle=0]{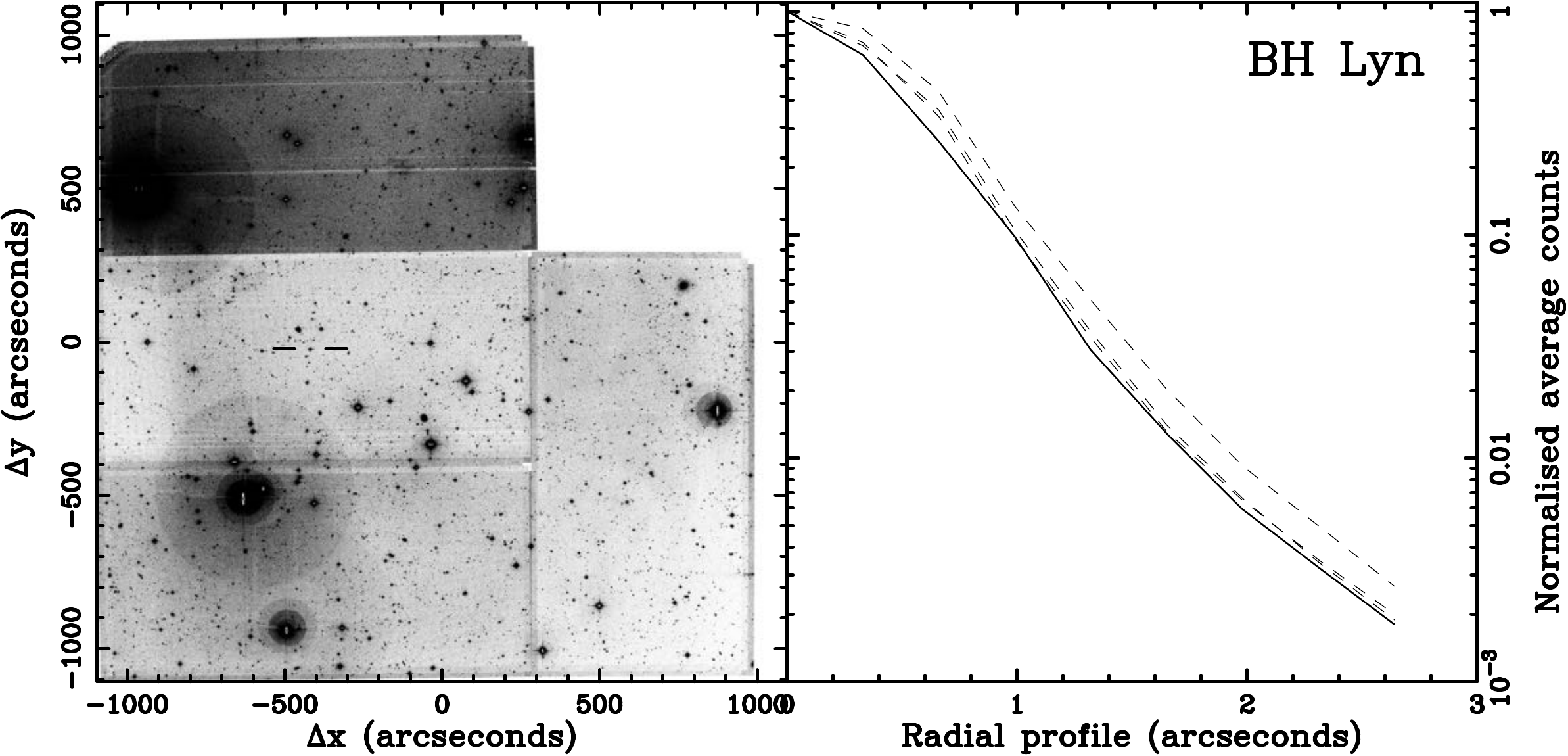}
  \vspace{10pt}
  \includegraphics[width=80mm,angle=0]{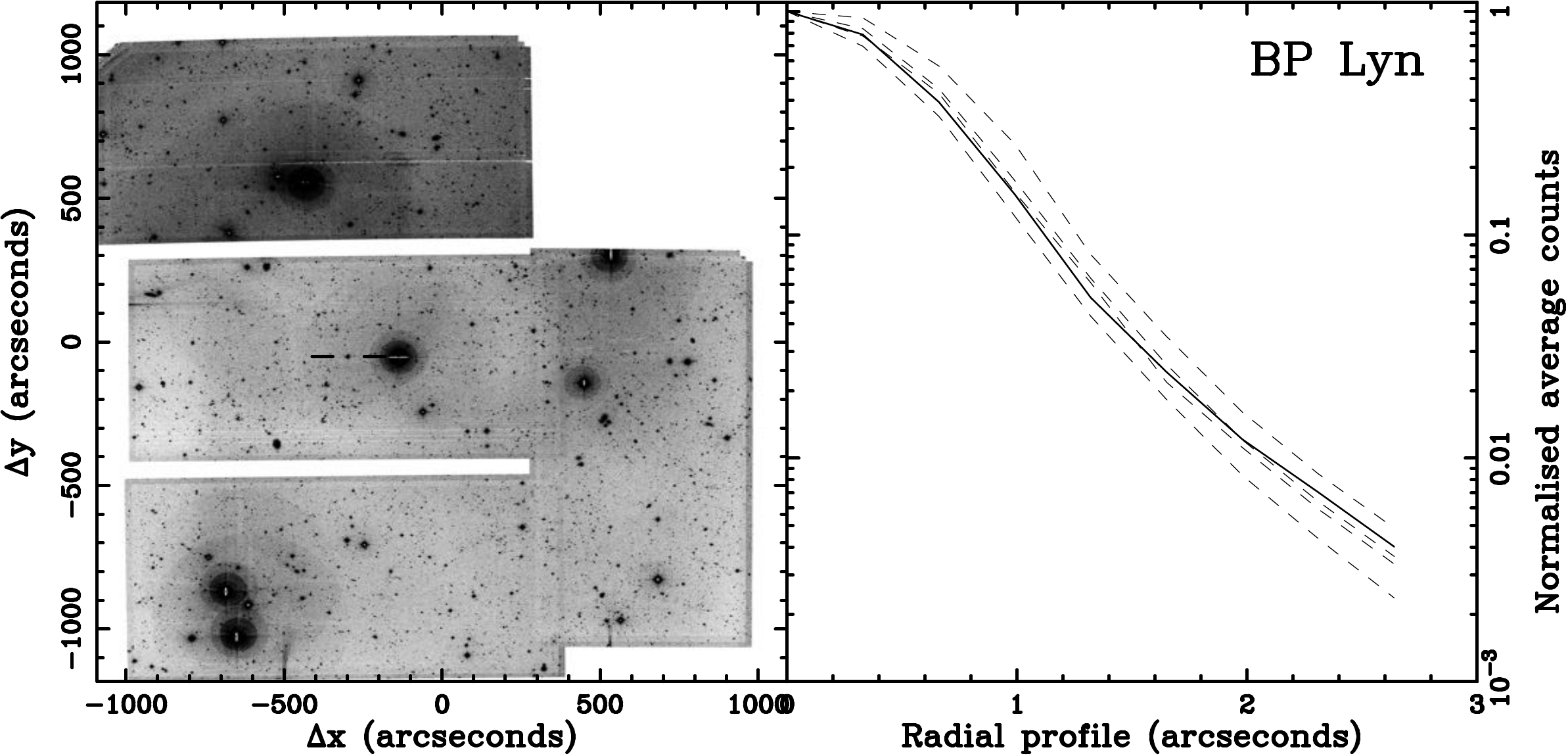}
\caption{See caption to Figure \ref{fig1} for details.}
 \label{fig5}
\end{figure}

\begin{figure}
\centering
  \includegraphics[width=80mm,angle=0]{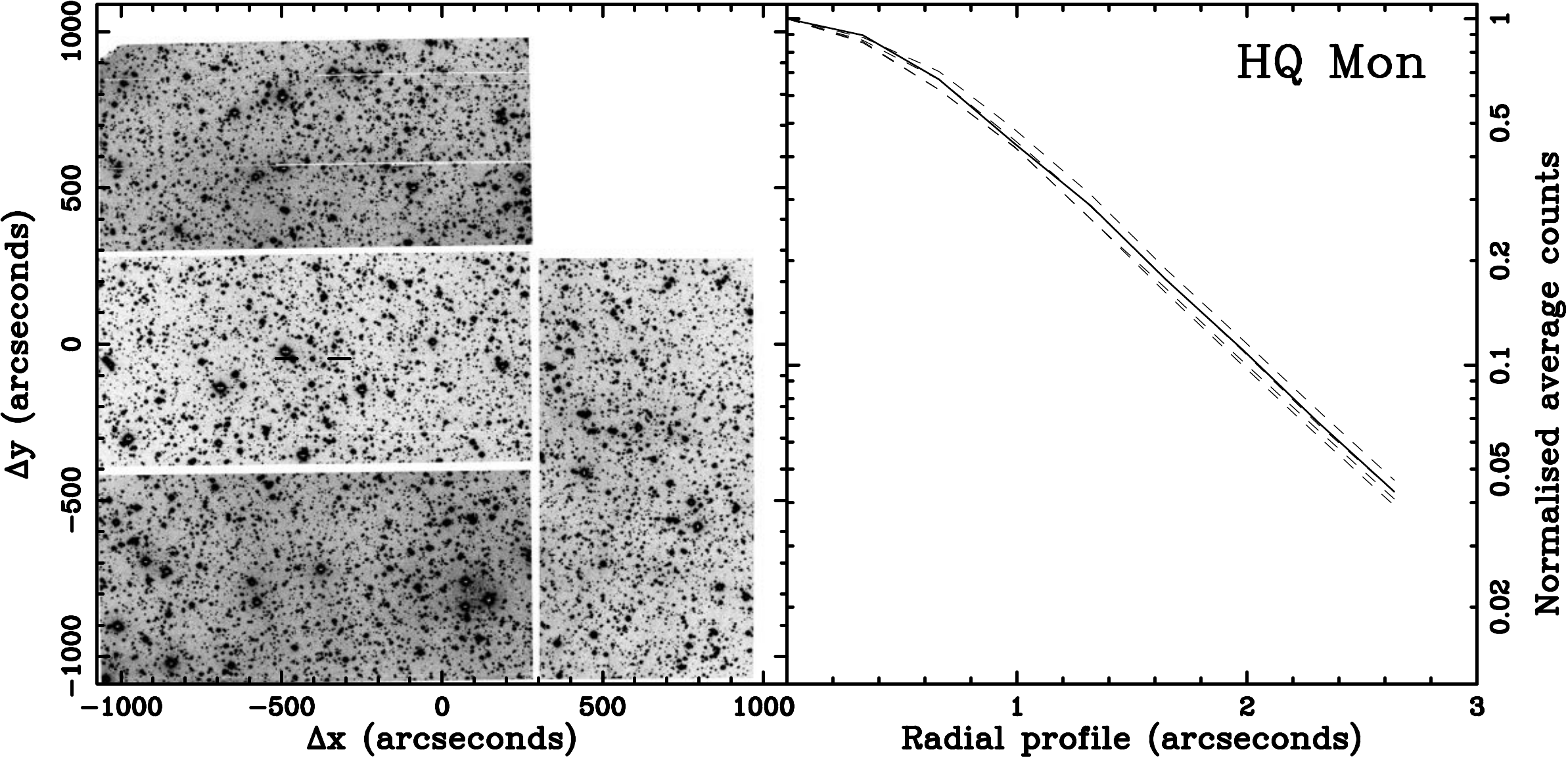}
  \vspace{10pt}
  \includegraphics[width=80mm,angle=0]{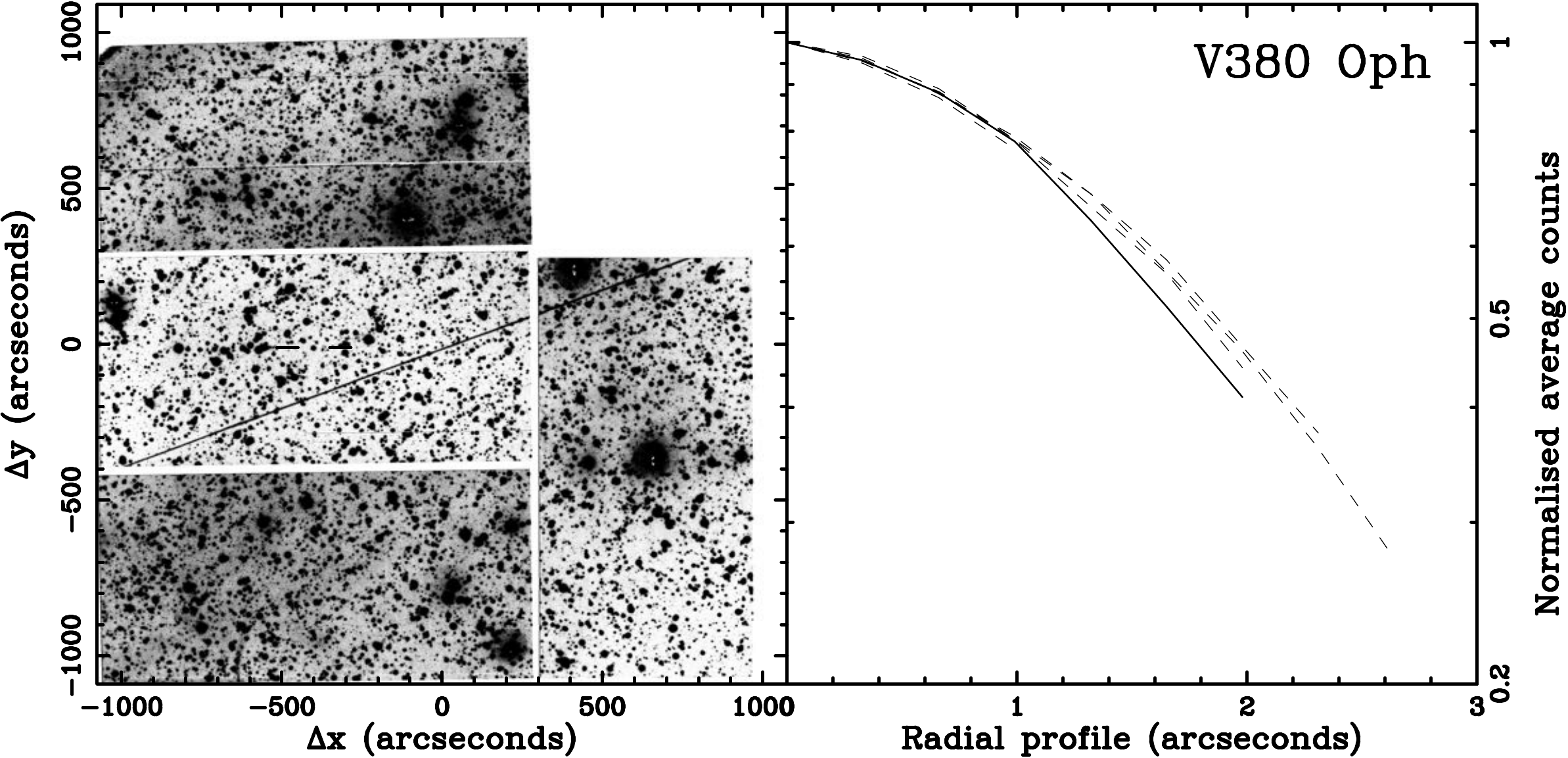}
  \vspace{10pt}
  \includegraphics[width=80mm,angle=0]{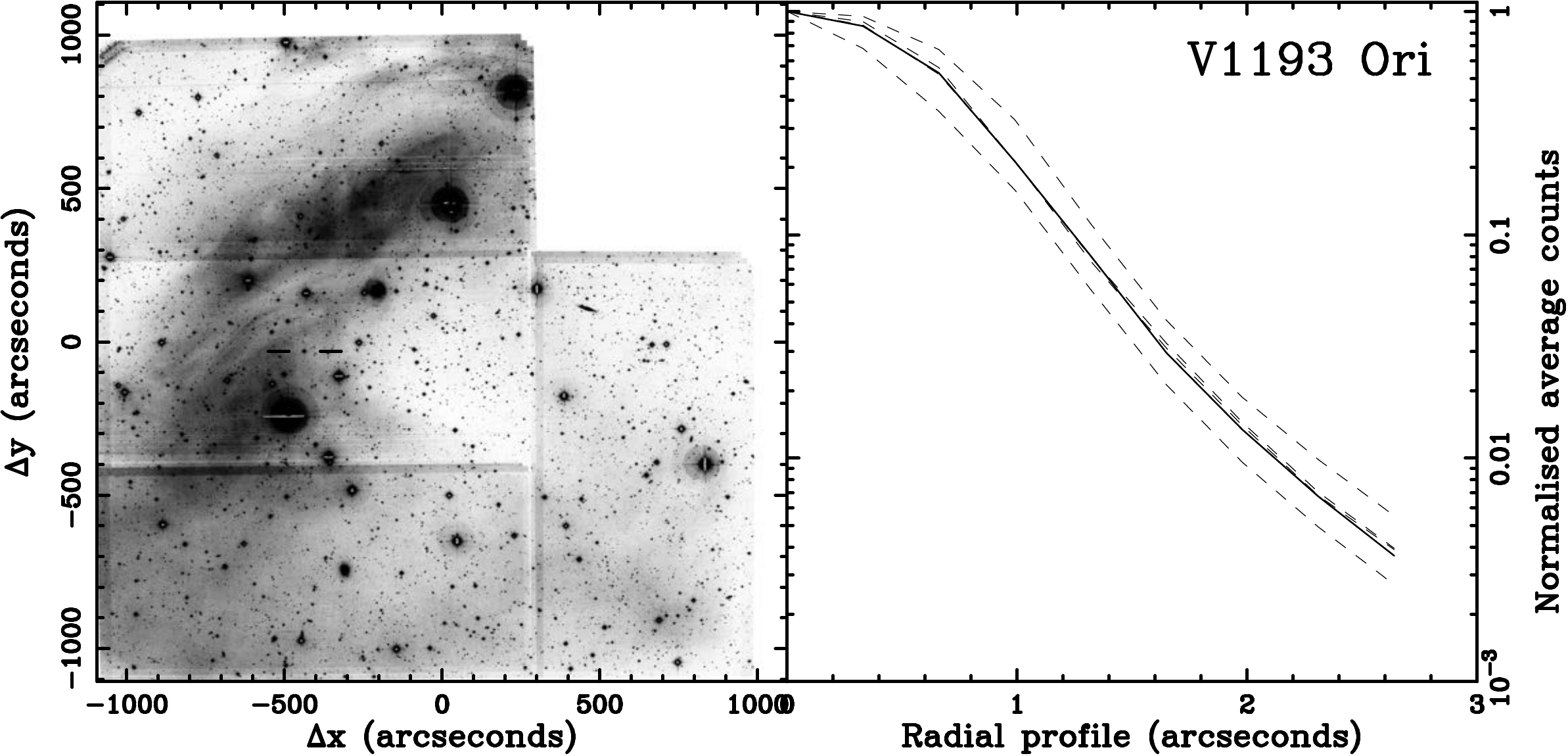}
  \vspace{10pt}
  \includegraphics[width=80mm,angle=0]{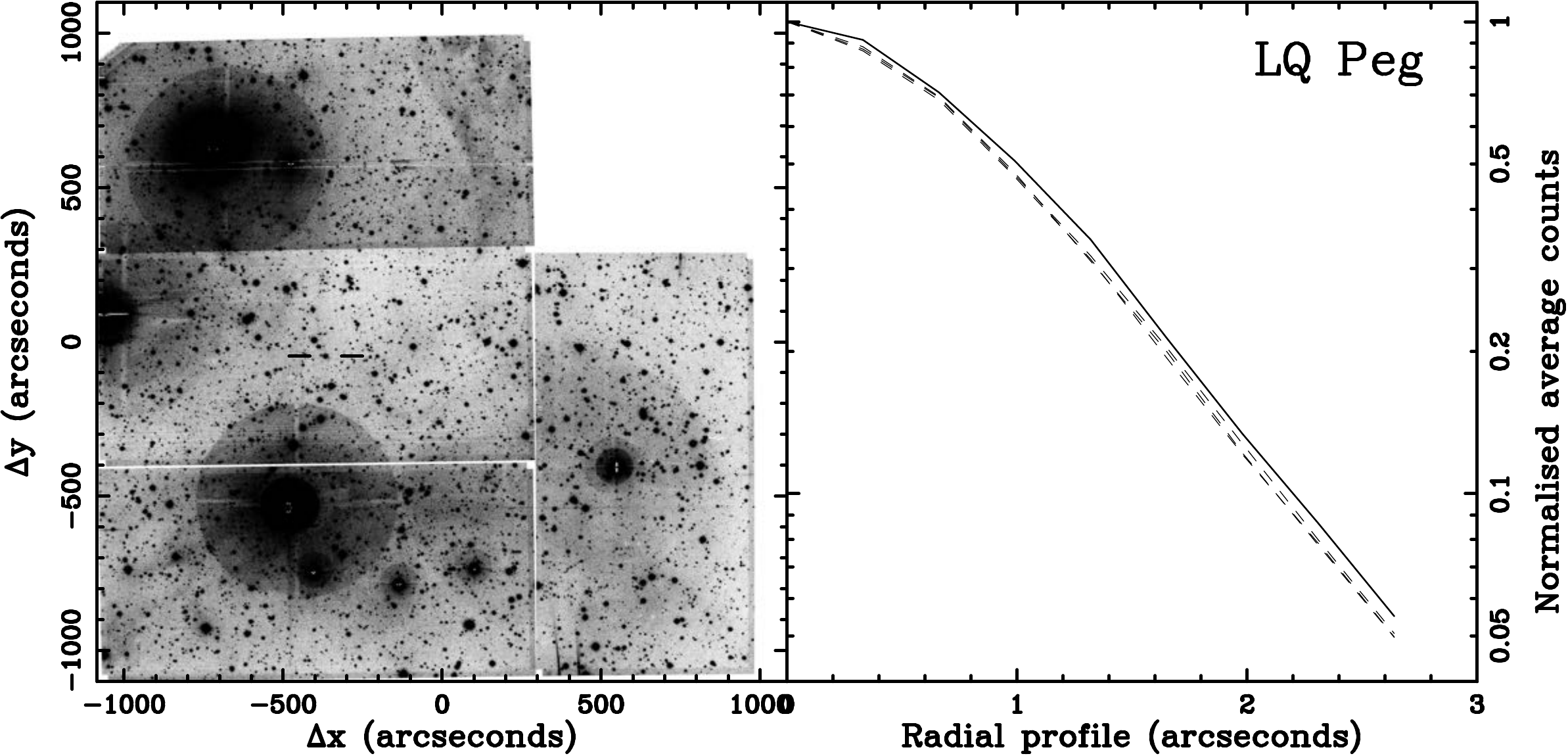}
 \vspace{10pt}
 \includegraphics[width=80mm,angle=0]{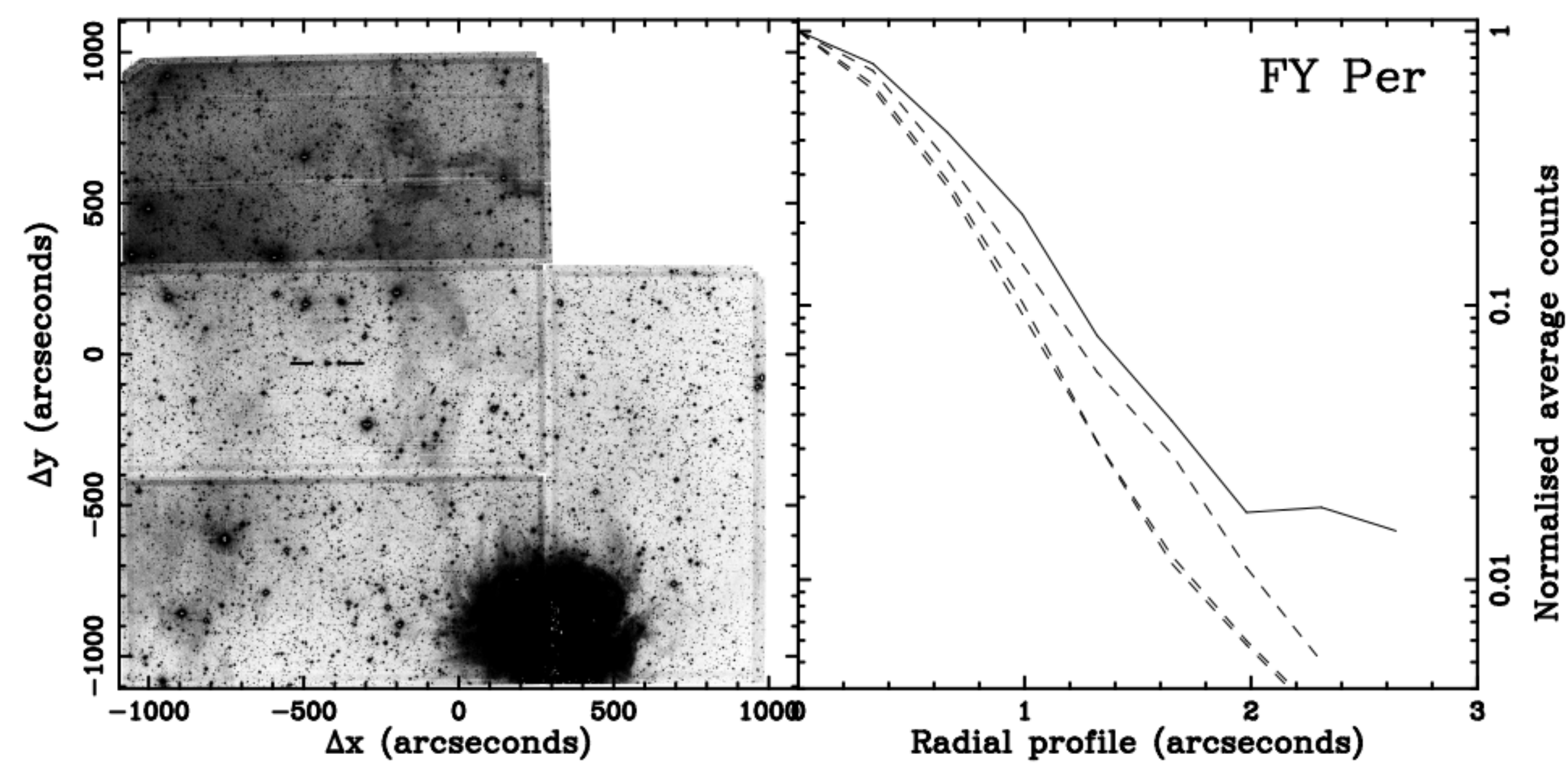}
\caption{See caption to Figure \ref{fig1} for details.}
 \label{fig6}
\end{figure}

\begin{figure}
\centering
  \vspace{10pt}
  \includegraphics[width=80mm,angle=0]{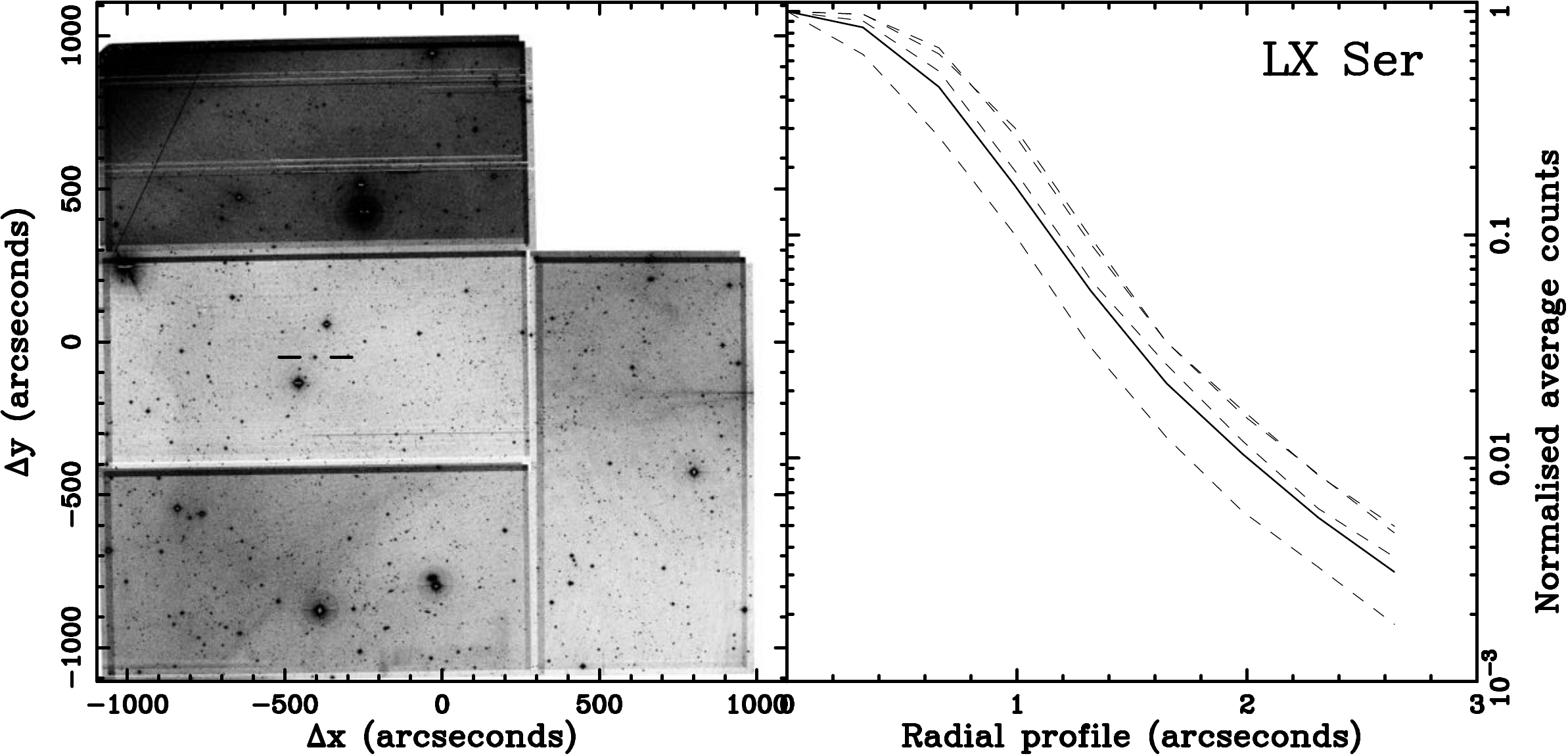}
  \vspace{10pt}
  \includegraphics[width=80mm,angle=0]{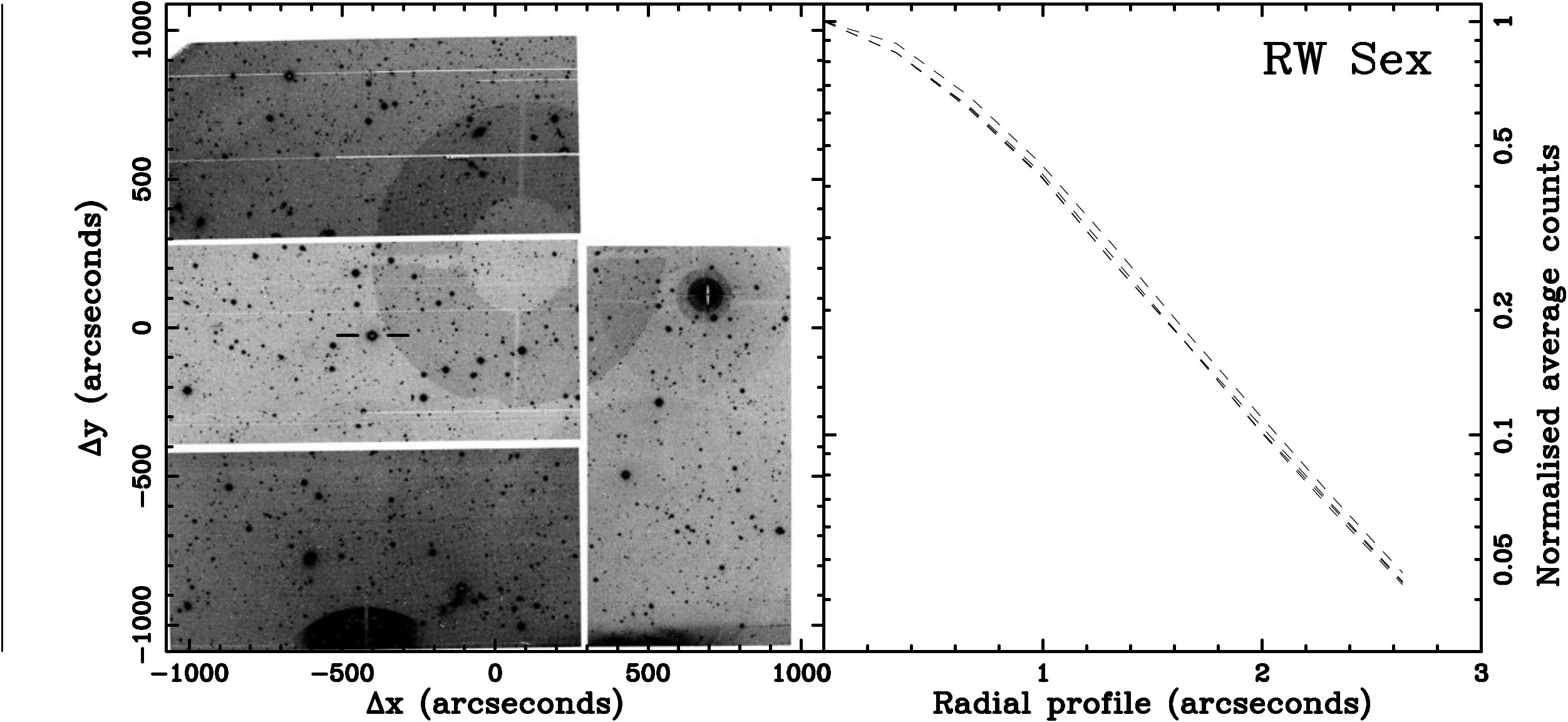}
  \vspace{10pt}
  \includegraphics[width=80mm,angle=0]{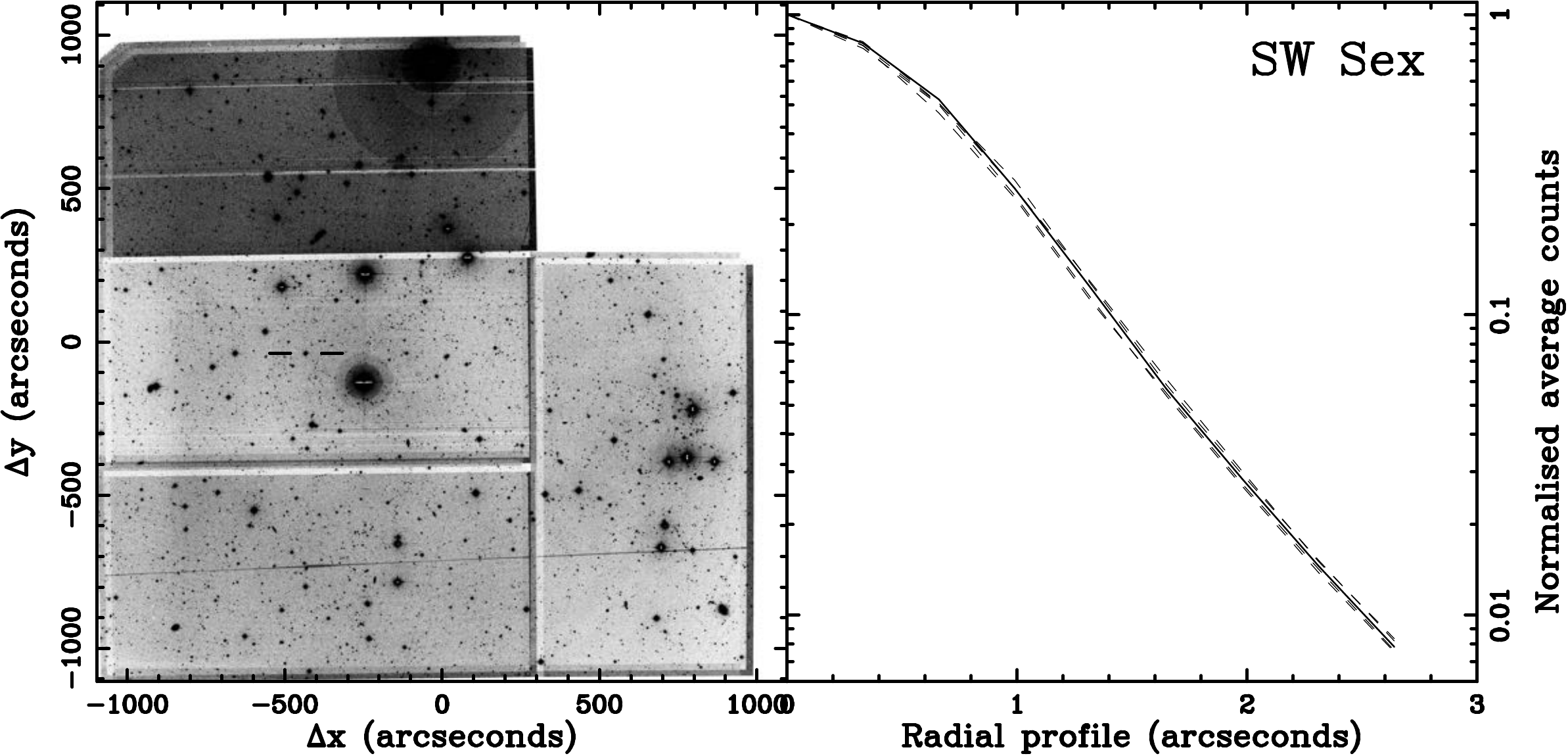}
  \vspace{10pt}
  \includegraphics[width=80mm,angle=0]{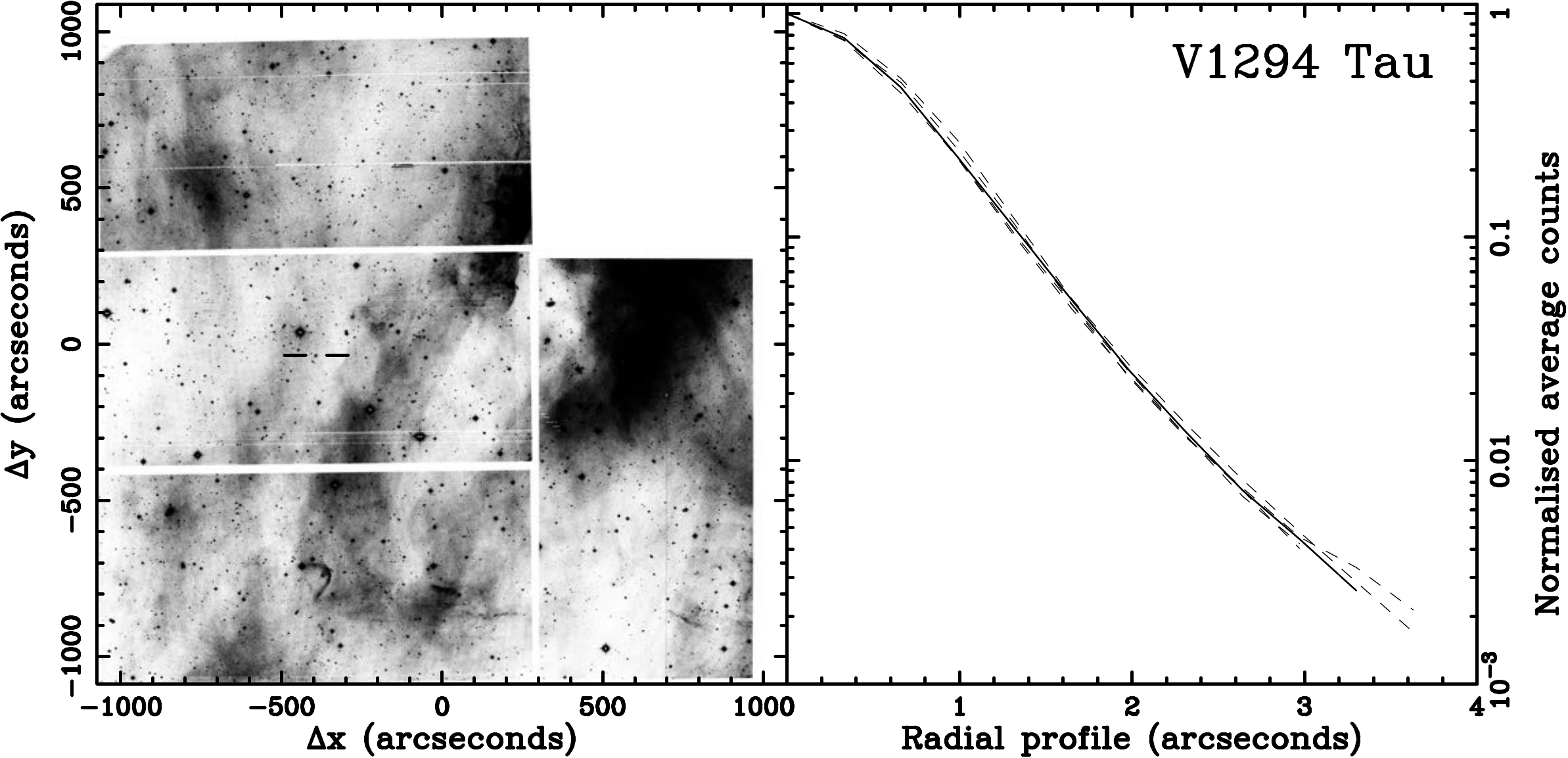}
  \vspace{10pt}
  \includegraphics[width=80mm,angle=0]{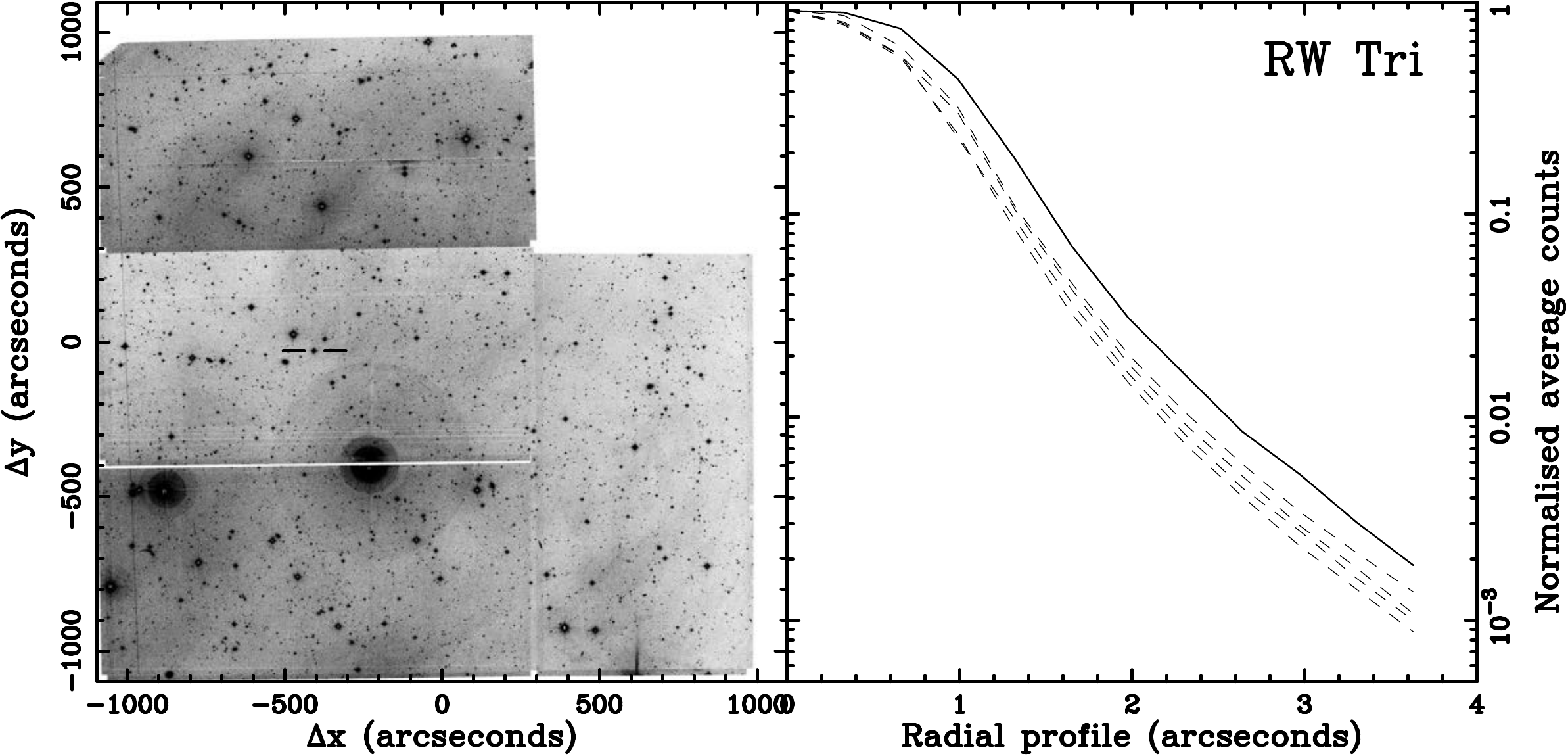}
  \caption{See caption to Figure \ref{fig1} for details.}
 \label{fig7}
\end{figure}

\begin{figure}
\centering
  \vspace{10pt}
  \includegraphics[width=80mm,angle=0]{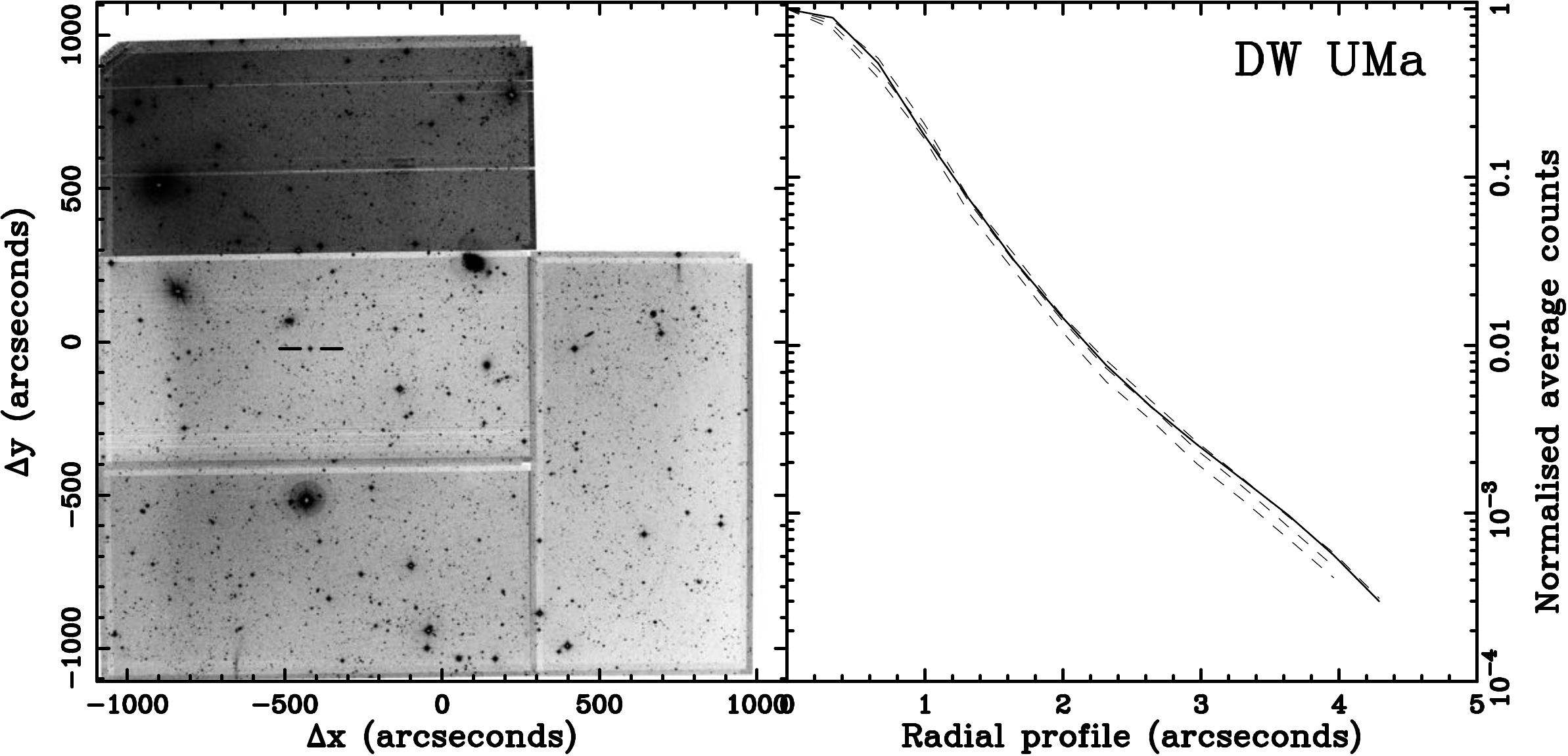}
  \vspace{10pt}
  \includegraphics[width=80mm,angle=0]{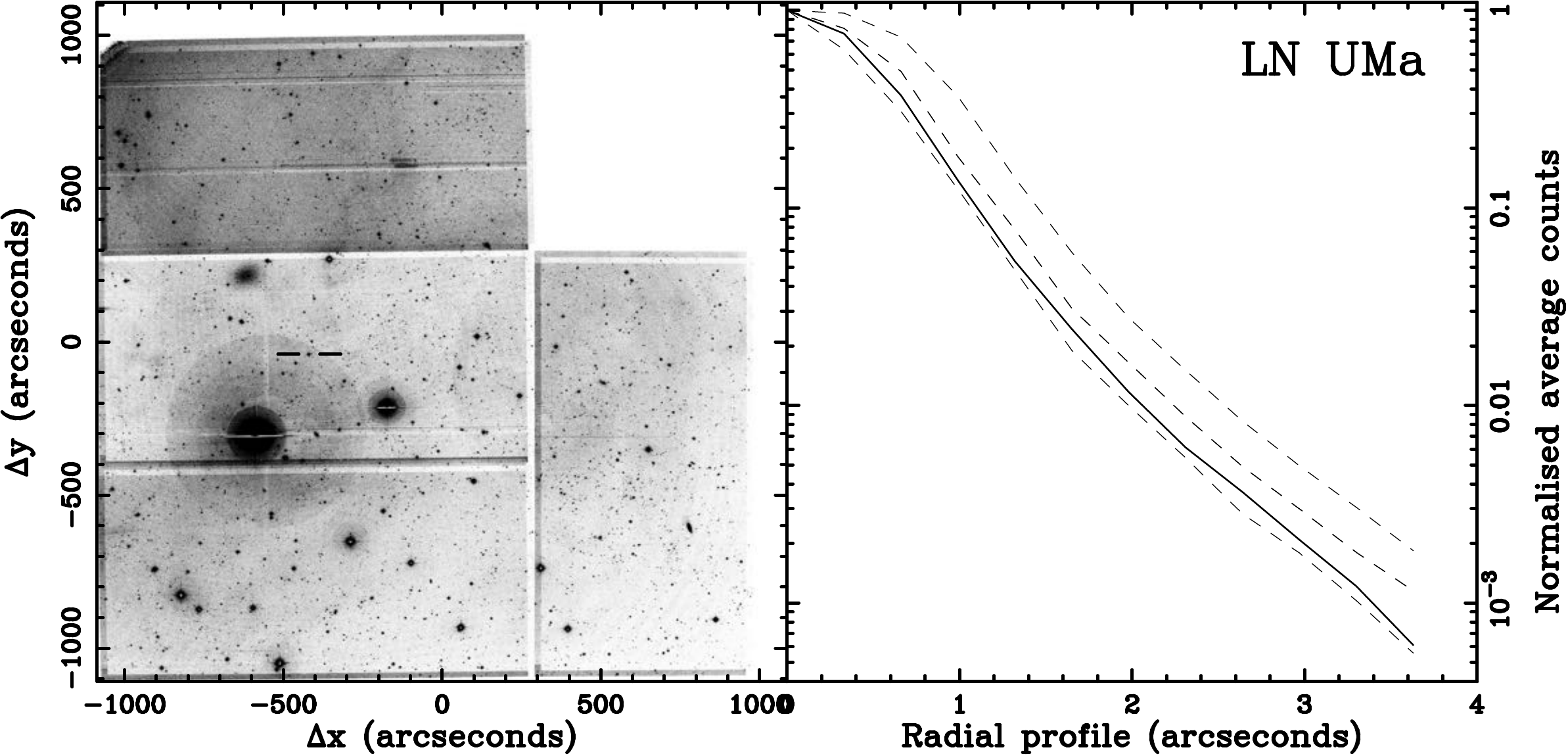}
  \vspace{10pt}
  \includegraphics[width=80mm,angle=0]{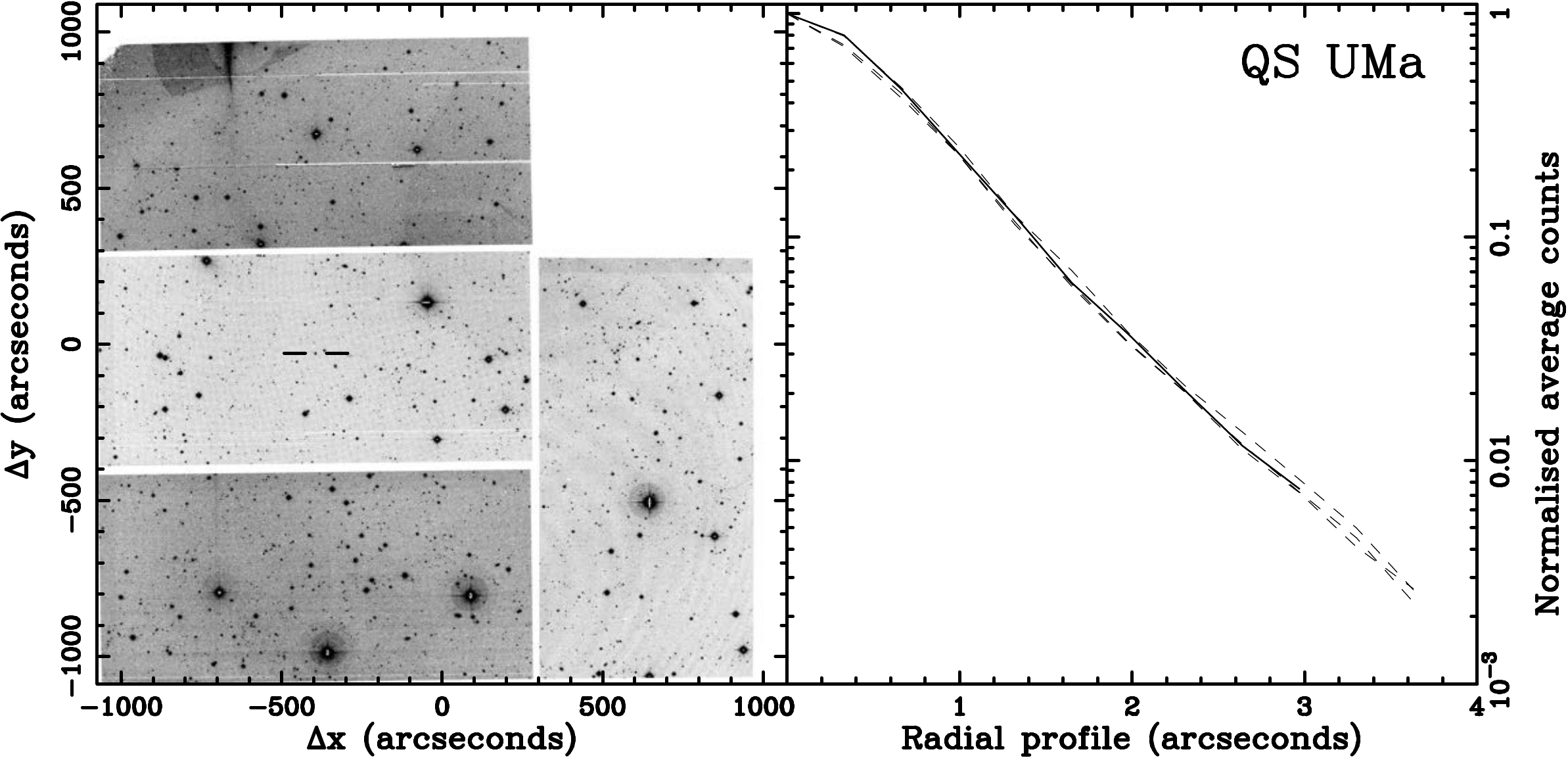}
  \vspace{10pt}
  \includegraphics[width=80mm,angle=0]{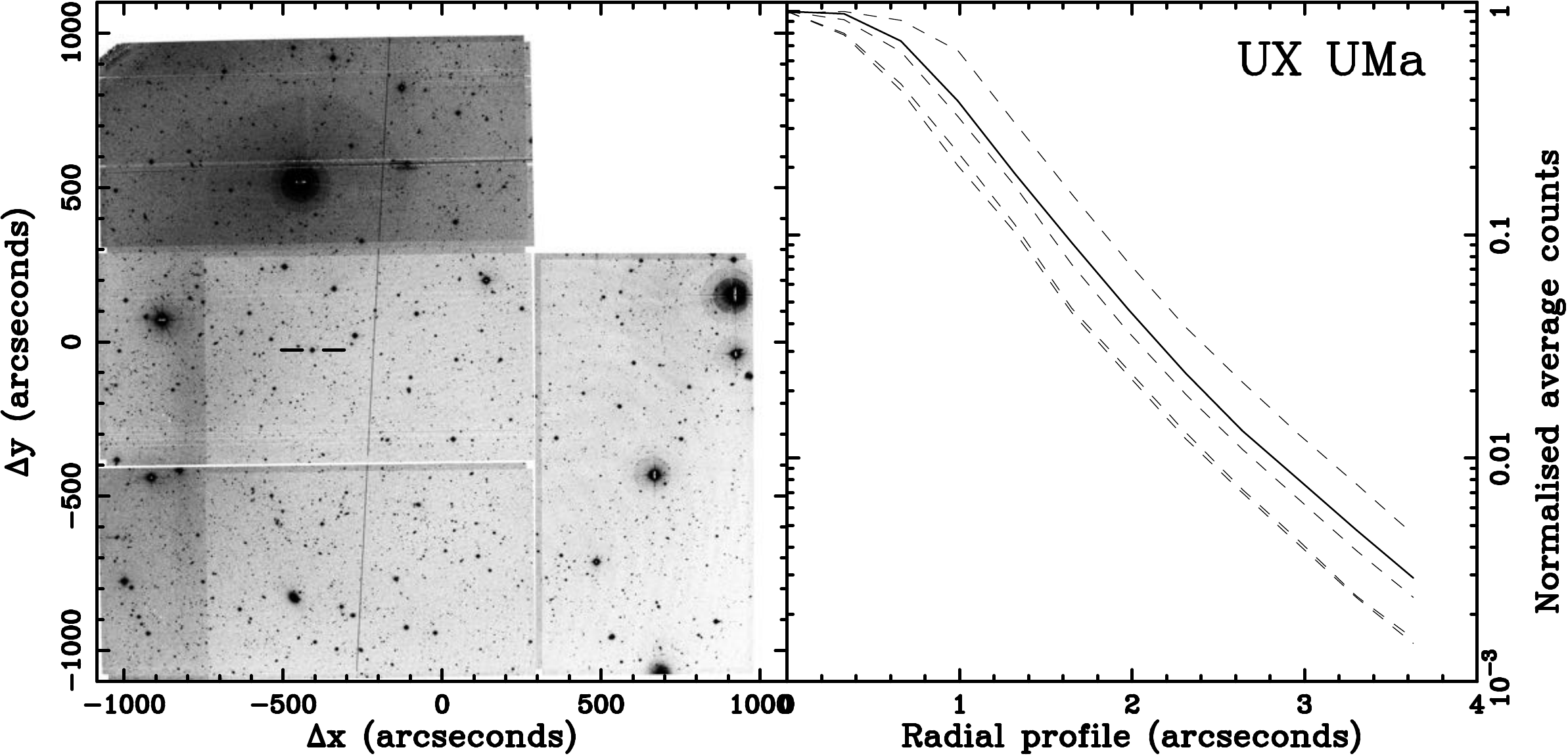}
  \vspace{10pt}
  \includegraphics[width=80mm,angle=0]{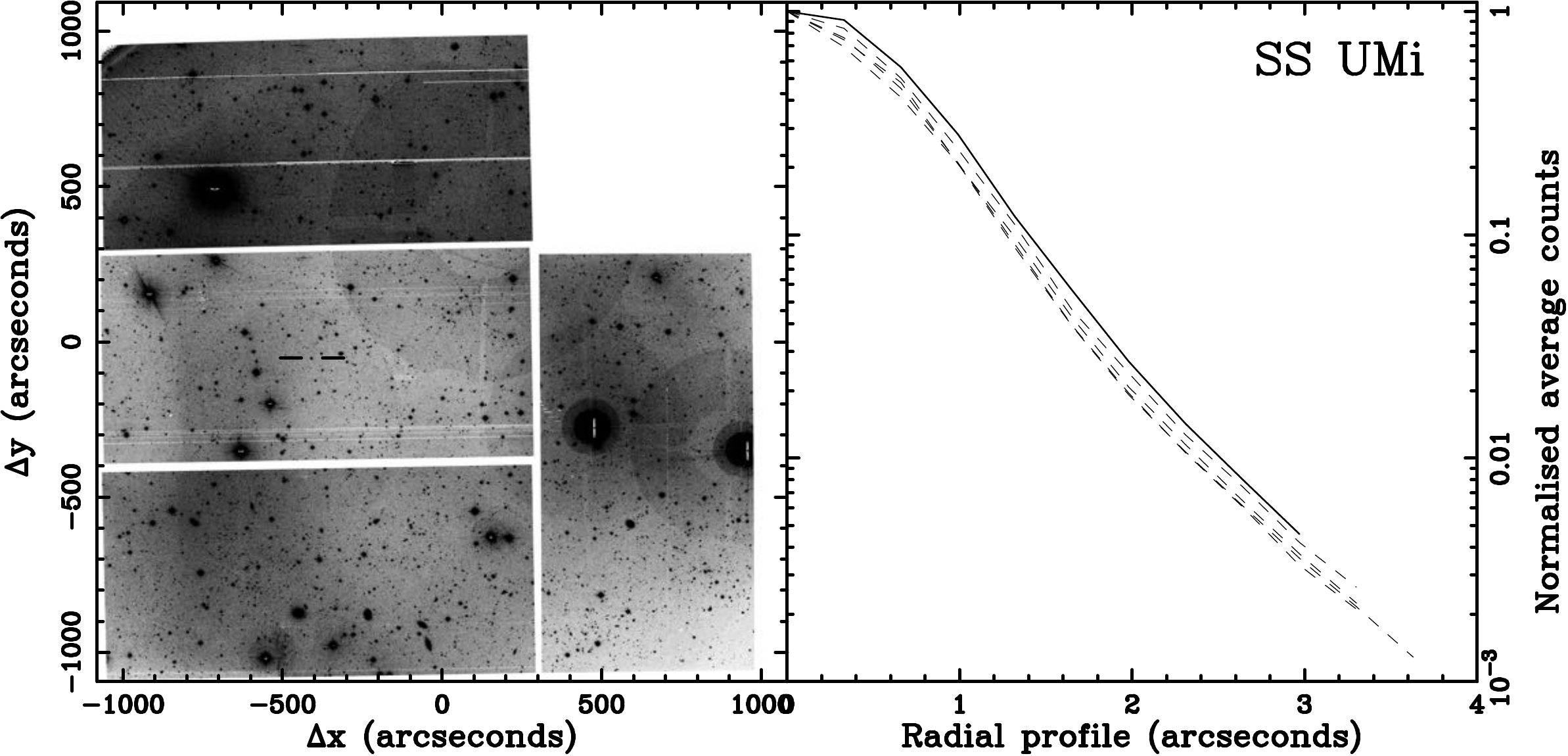}
 \caption{See caption to Figure \ref{fig1} for details.}
 \label{fig8}
\end{figure}
  
\begin{figure}
\centering
  \vspace{10pt}
  \includegraphics[width=80mm,angle=0]{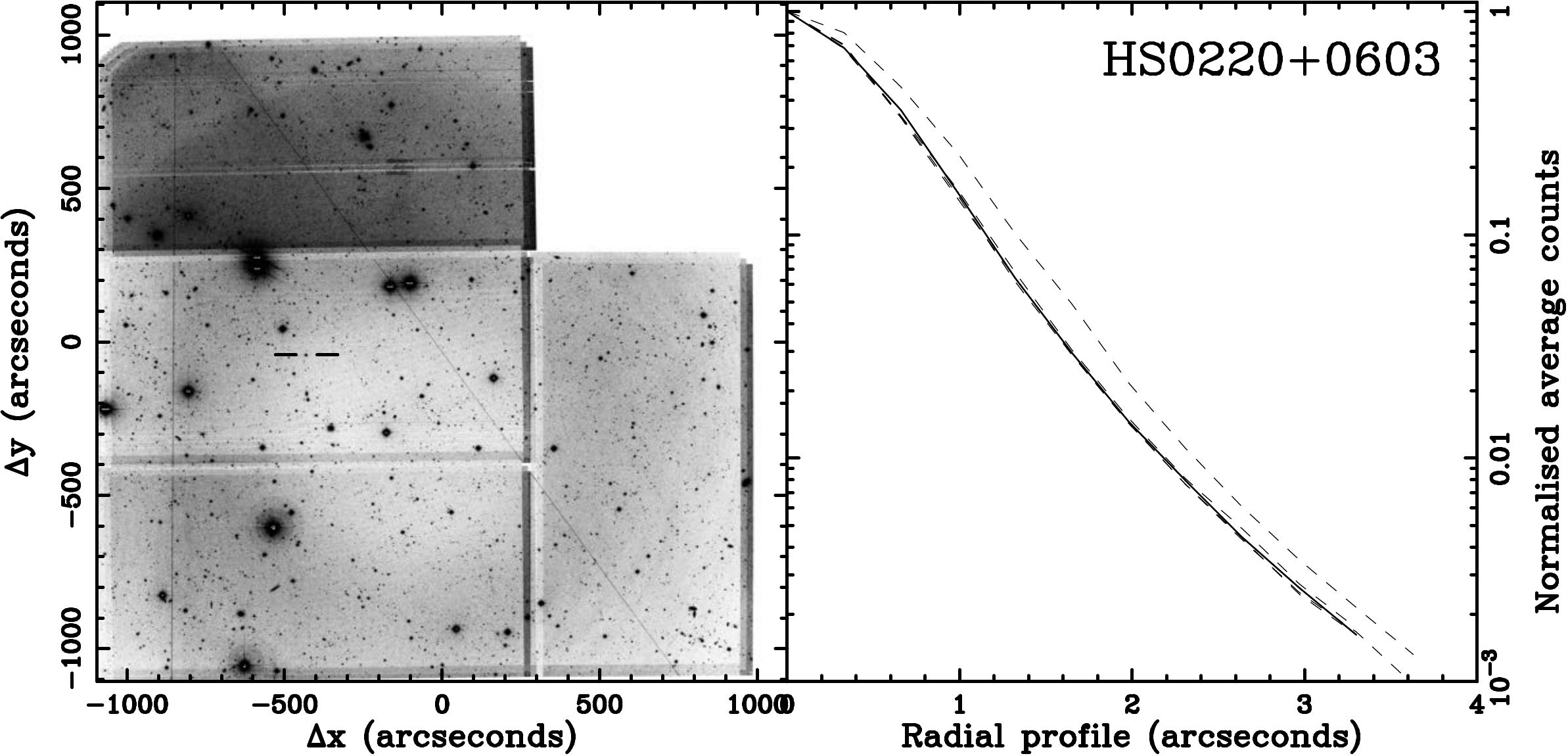}
  \vspace{10pt}
  \includegraphics[width=80mm,angle=0]{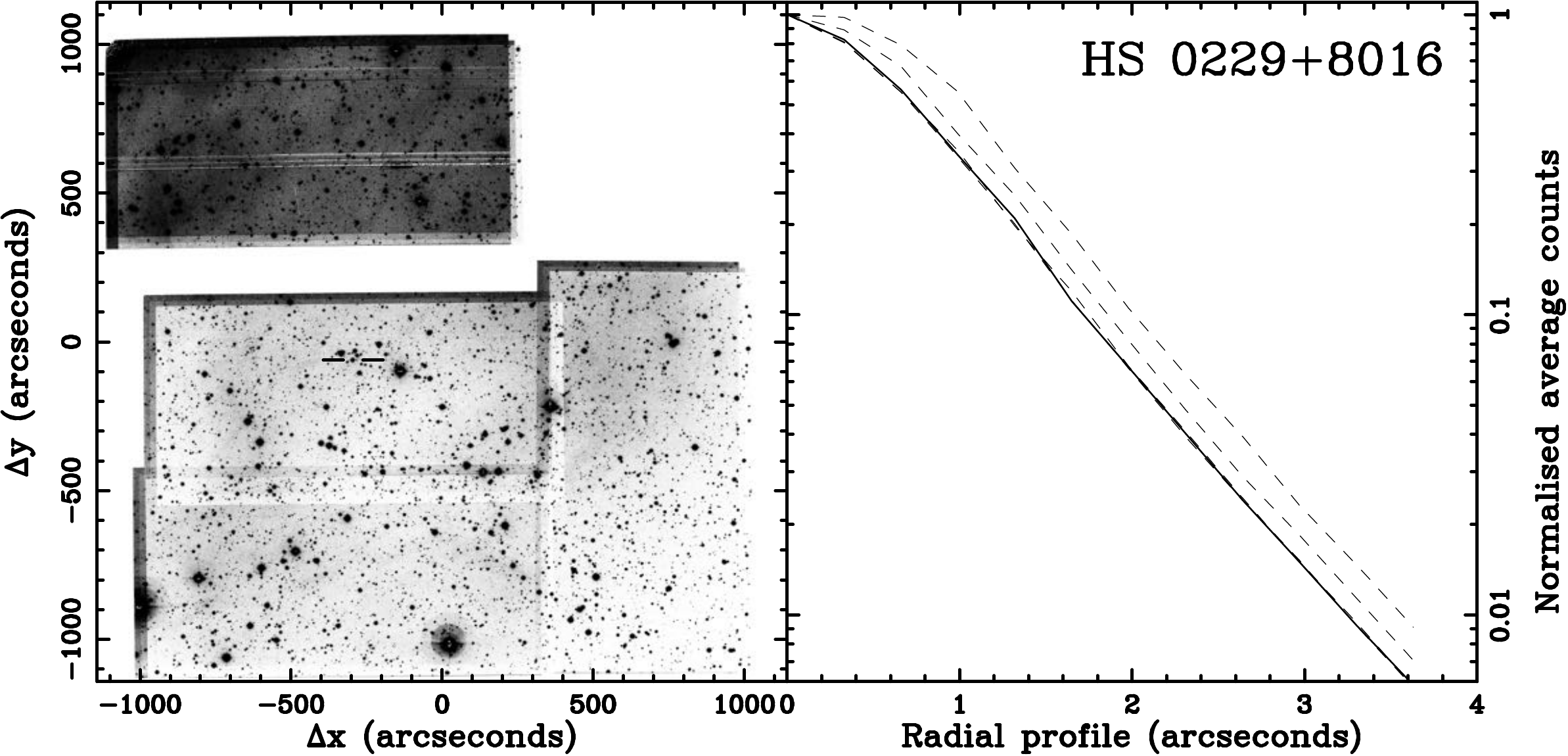}
  \vspace{10pt}
  \includegraphics[width=80mm,angle=0]{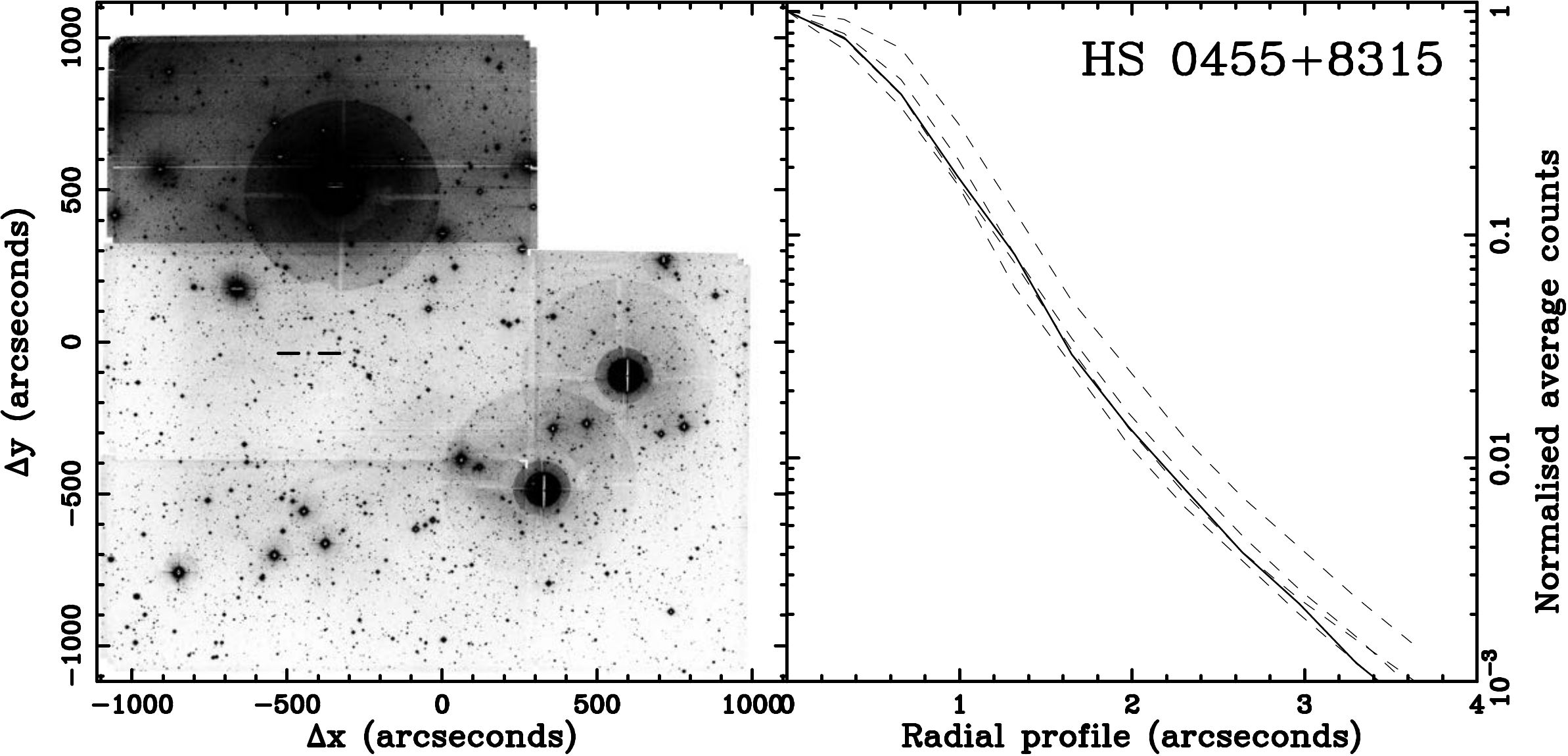}
  \vspace{10pt}
  \includegraphics[width=80mm,angle=0]{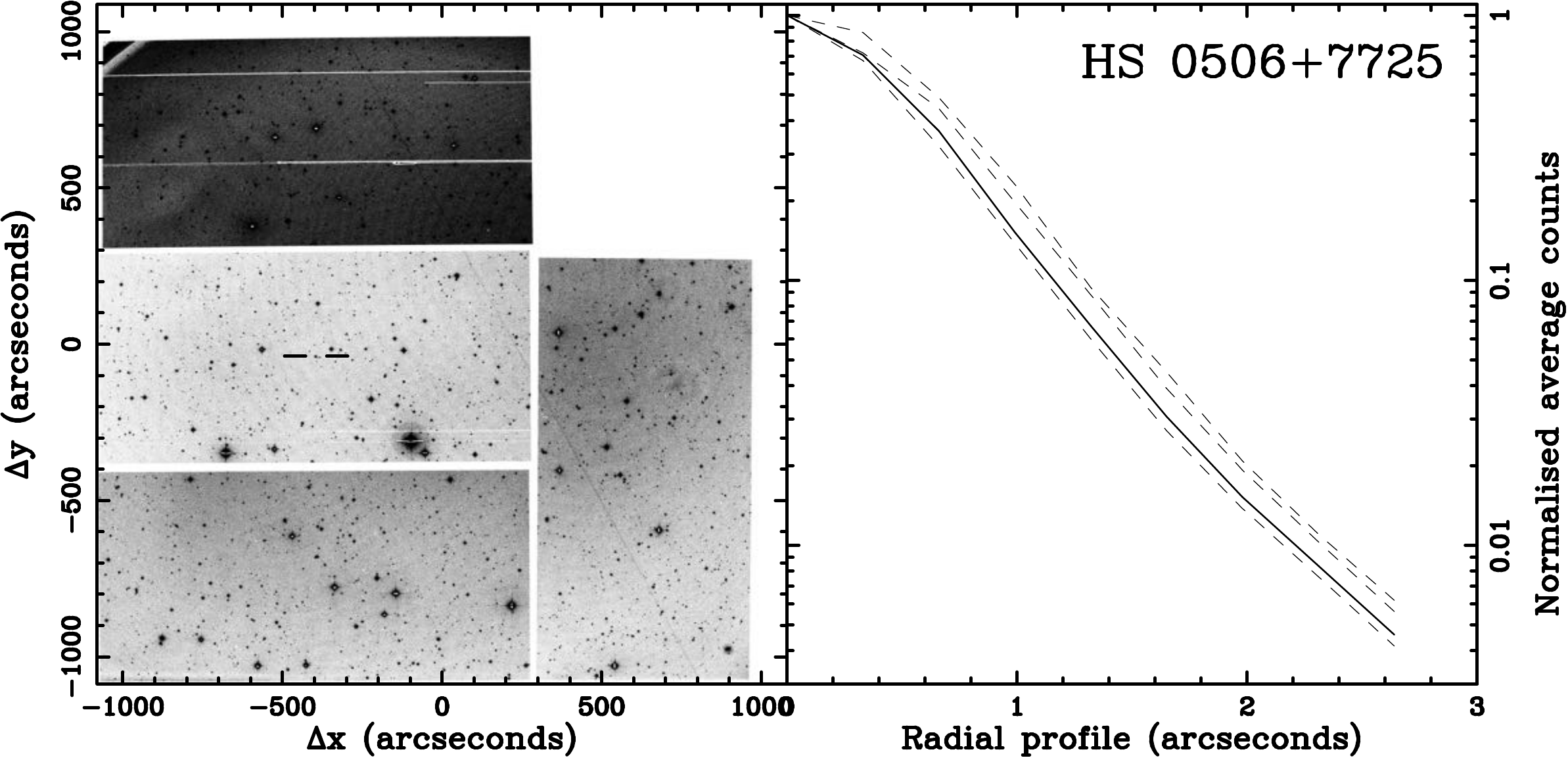}
  \vspace{10pt}
  \includegraphics[width=80mm,angle=0]{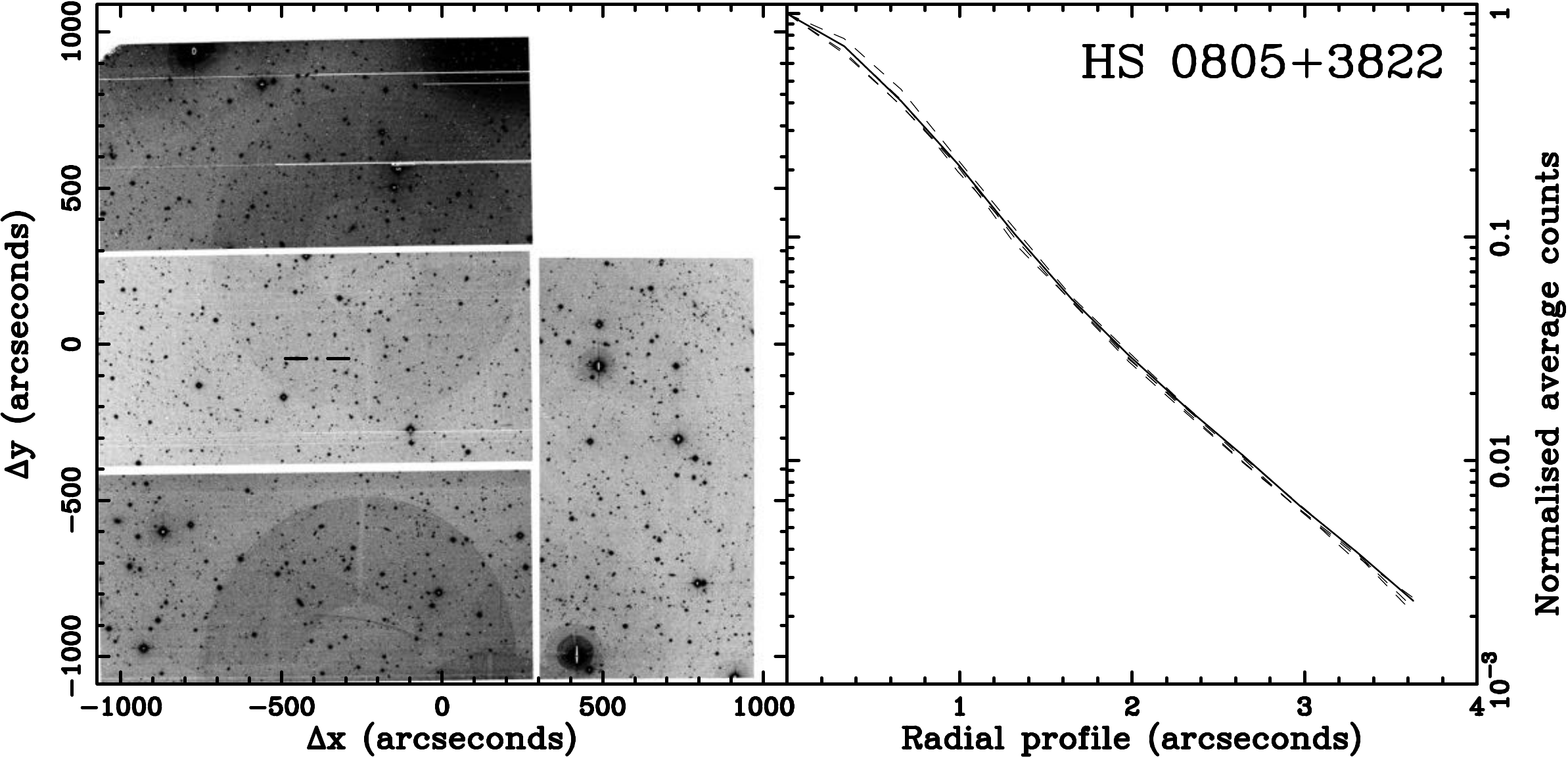}
 \caption{See caption to Figure \ref{fig1} for details.}
 \label{fig10}
\end{figure}
  
\begin{figure}
  \centering
  \vspace{10pt}
  \includegraphics[width=80mm,angle=0]{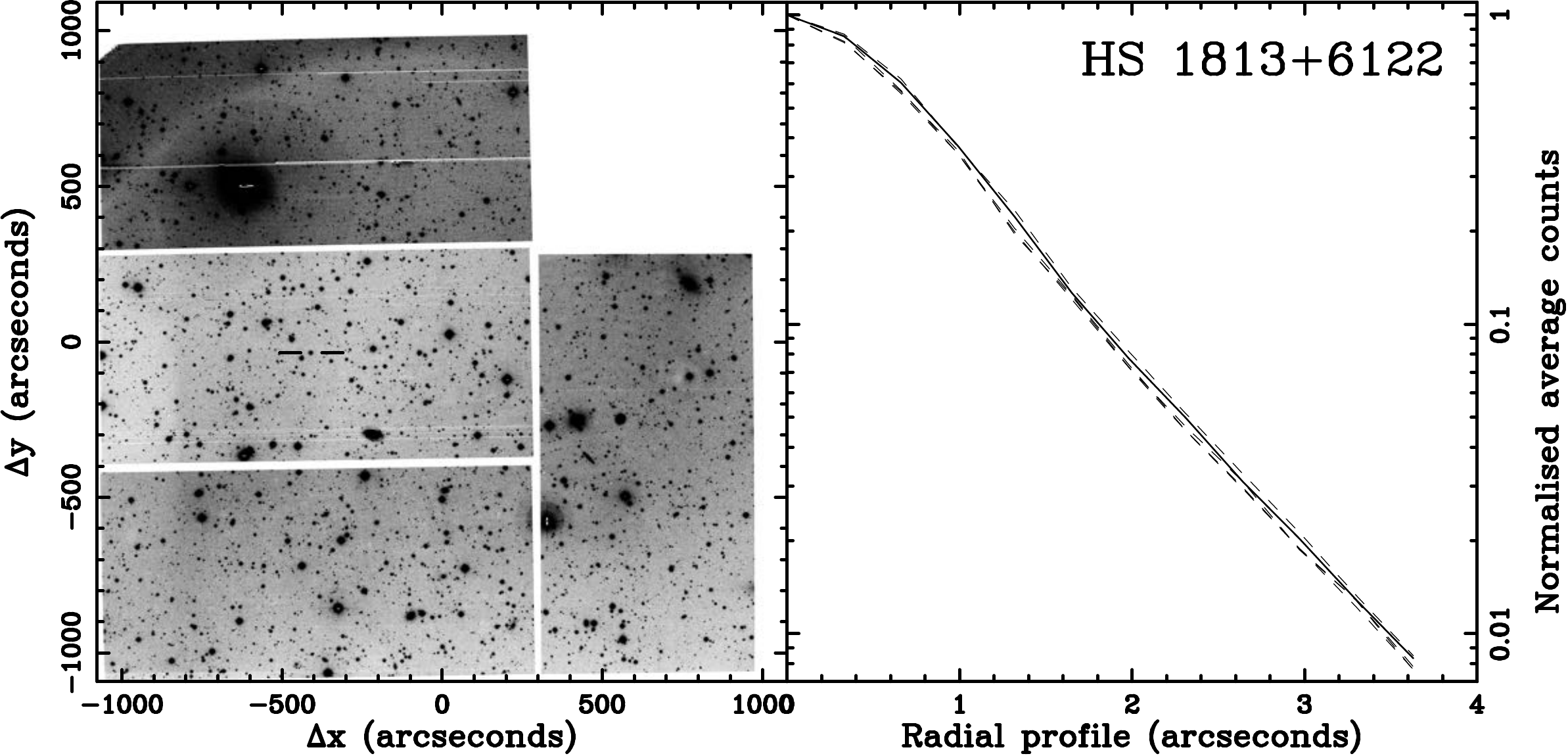}
  \vspace{10pt}
  \includegraphics[width=80mm,angle=0]{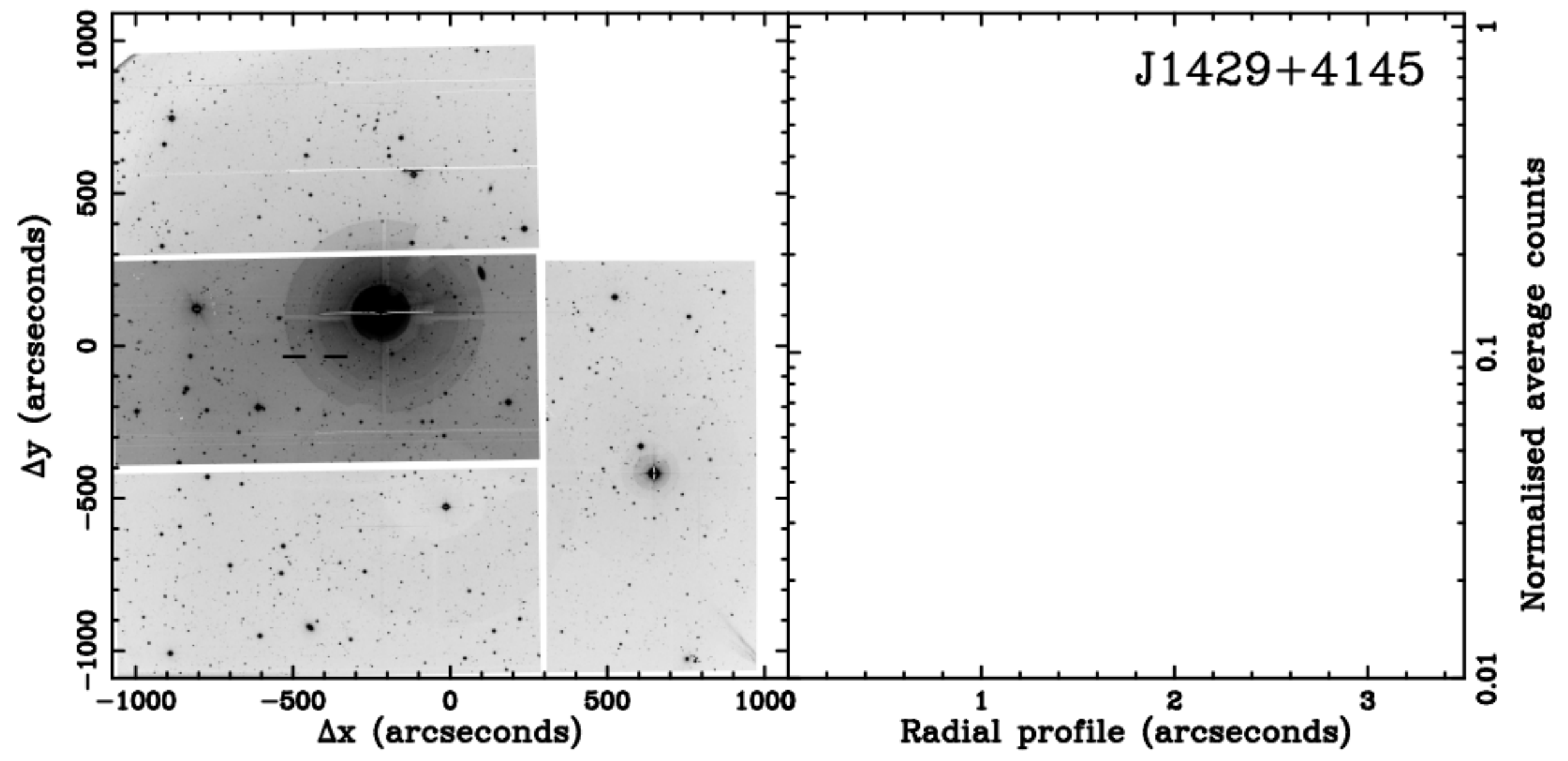}
  \vspace{10pt}
  \includegraphics[width=80mm,angle=0]{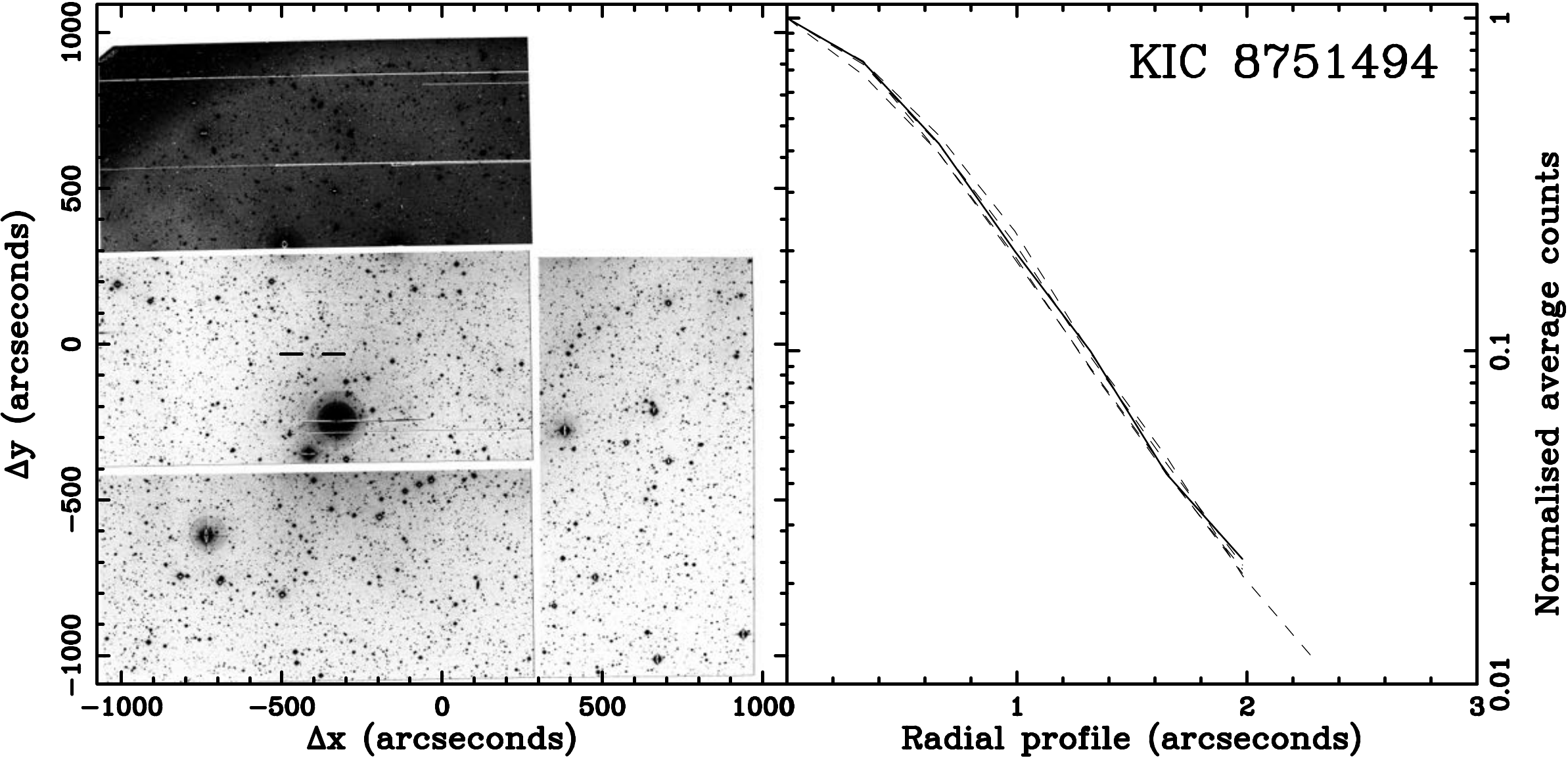}
  \vspace{10pt}
  \includegraphics[width=80mm,angle=0]{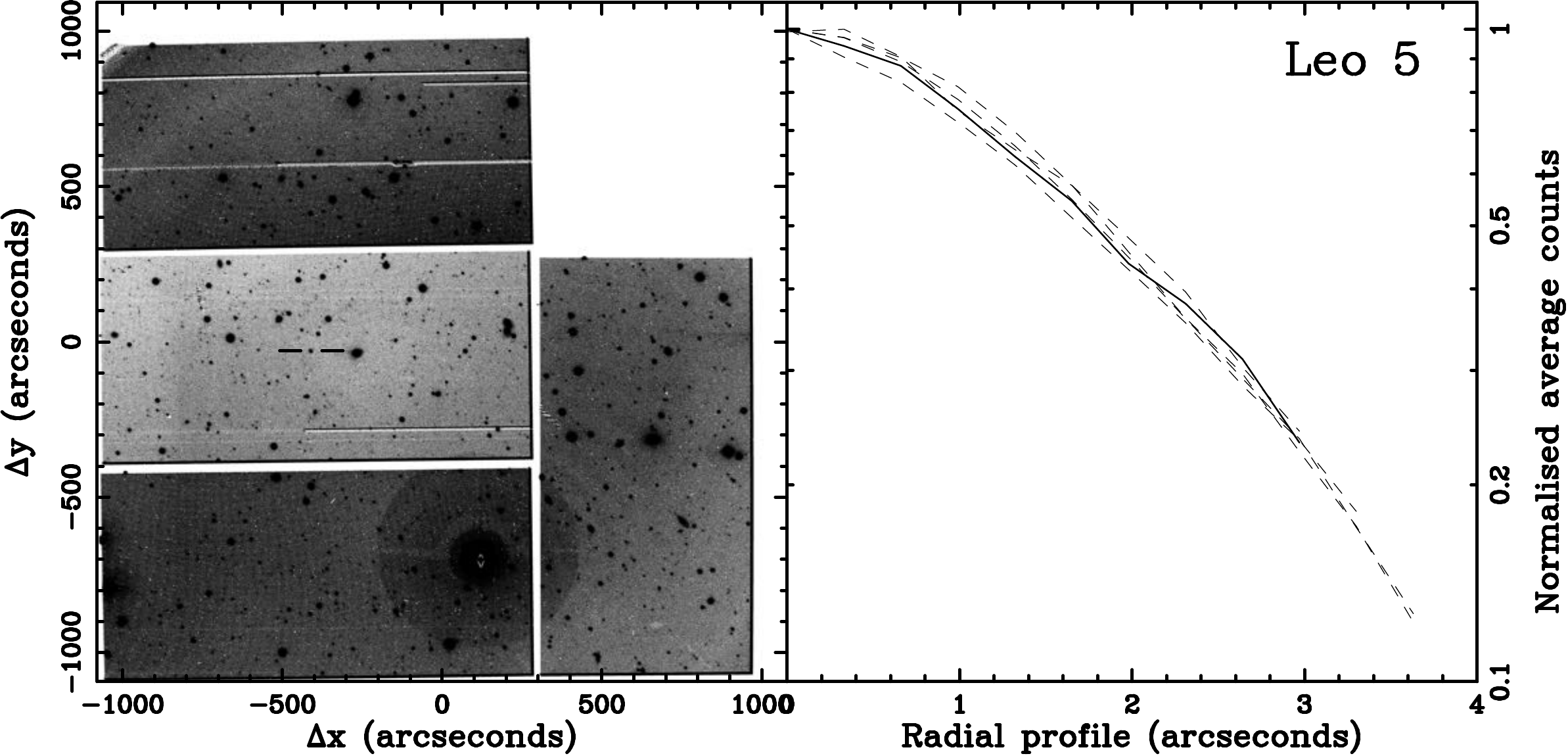}
 \vspace{10pt}
 \includegraphics[width=80mm,angle=0]{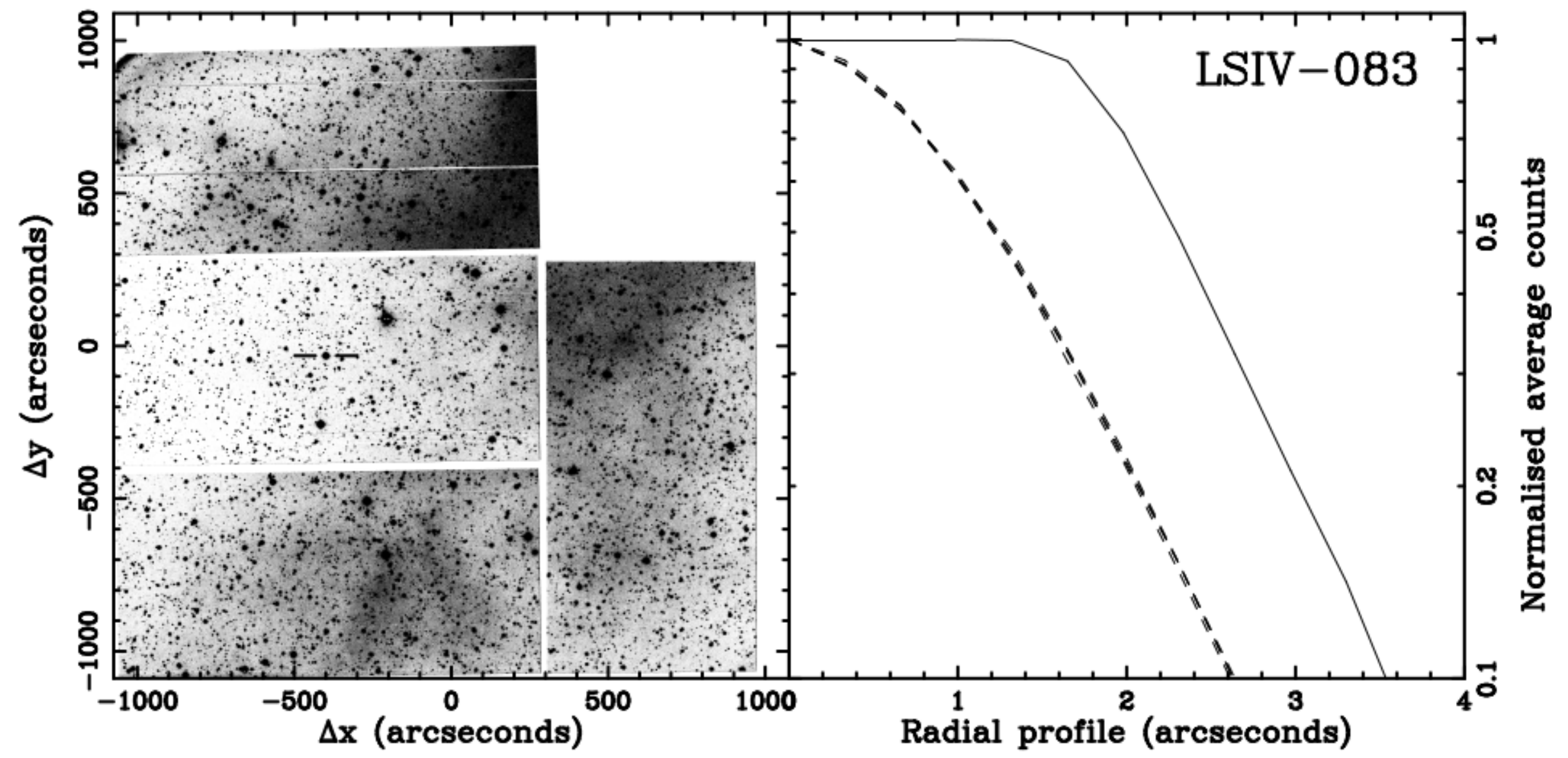}
  \caption{See caption to Figure \ref{fig1} for details.}
 \label{fig11}
\end{figure}

\begin{figure}
\centering
  \includegraphics[width=80mm,angle=0]{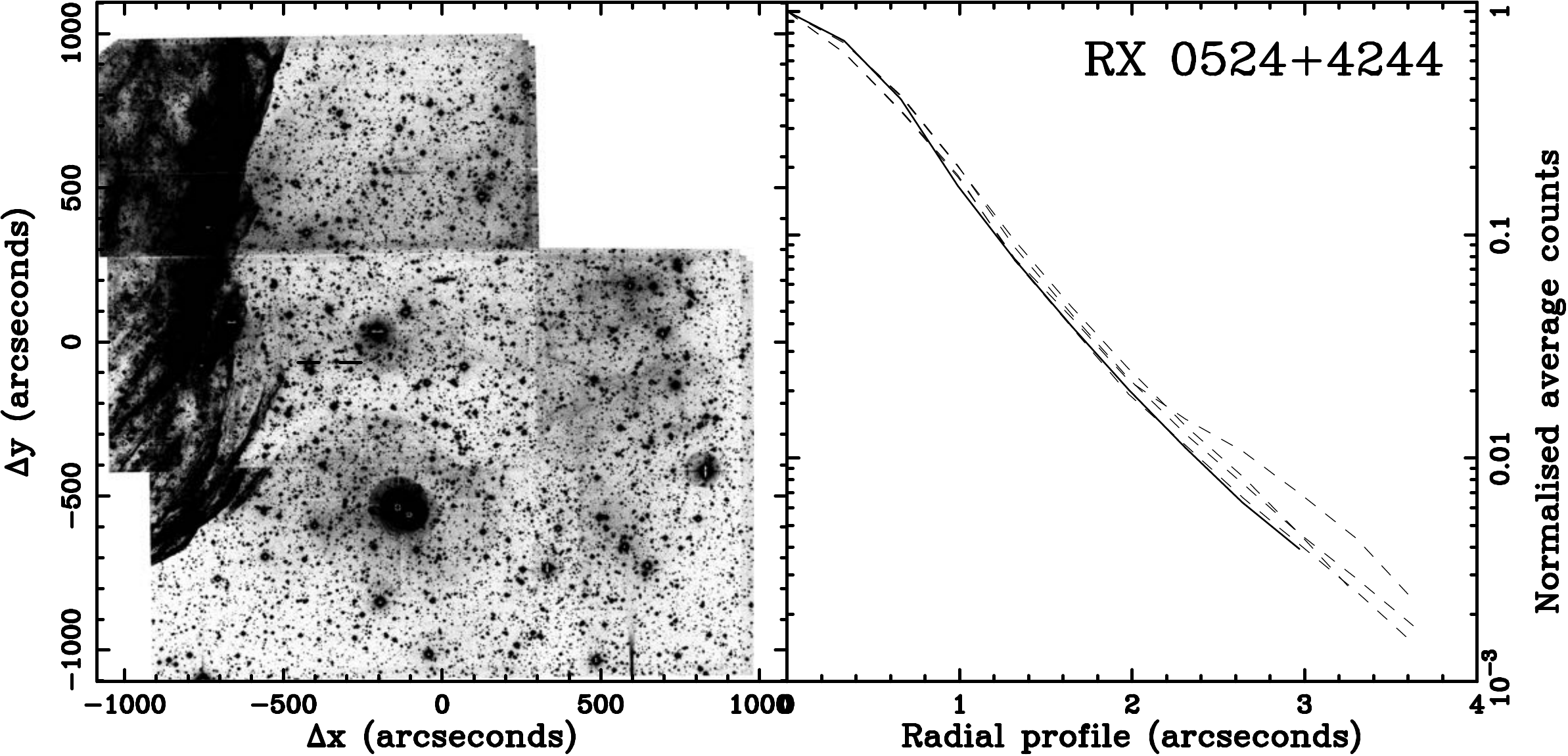}
  \vspace{10pt}
\caption{See caption to Figure \ref{fig1} for details.}
 \label{fig12}
\end{figure}

\label{lastpage}

\end{document}